# Topology optimization of anisotropic elastic metamaterial with broadband double-negative index

Hao-Wen Dong[1, 2], Sheng-Dong Zhao[1, 2], Yue-Sheng Wang[1,*], and Chuanzeng Zhang[2,*]

[1]Institute of Engineering Mechanics, Beijing Jiaotong University, Beijing 100044, China

[2]Department of Civil Engineering, University of Siegen, D-57068 Siegen, Germany

**Abstract:** As the counterpart of electromagnetic and acoustic metamaterials, elastic metamaterials, artificial periodic composite materials, also offer the ultimate possibility to manipulate elastic wave propagation in the subwavelength scale through different mechanisms. Aiming at the promising superlensing for the medical ultrasonic and detection, the double-negative metamaterials which possess the negative mass density and elastic modulus simultaneously can be acted as the ideal superlens for breaking the diffraction limit. In this paper, we use topology optimization to design the two-dimensional single-phase anisotropic elastic metamaterials with broadband double-negative indices and numerically demonstrate the superlensing at the deep-subwavelength scale. We also discuss the impact of several parameters adopted in the objective function and constraints on the optimized results. Unlike all previous reported mechanisms, our optimized structures exhibit the new quadrupolar or multipolar resonances for the negative mass density, negative longitudinal and shear moduli. In addition, we observe the negative refraction of transverse waves in a single-phase material. Most structures can serve as the anisotropic zero-index metamaterials for the longitudinal or transverse wave at a certain frequency. The cloaking effect is demonstrated for both the longitudinal and transverse waves. Moreover, with the particular constraints in optimization, we design a super-anisotropic metamaterial exhibiting the double-negative and hyperbolic dispersions along two principal directions, respectively. Our optimization work provides a robust computational approach to negative index engineering in elastic metamaterials and guides design of other kinds of metamaterials, including the electromagnetic and acoustic metamaterials. The unusual properties of our optimized structures are likely to inspire new ideas and novel applications including the low-frequency vibration attenuation, flat lens and ultrasonography for elastic waves in the future.

**Keywords:**

Elastic metamaterial; Topology optimization; Double-negative index; Anisotropy; Superlensing; Cloaking effect; Hyperbolic dispersion

* Corresponding author (Y. S. Wang) Email: yswang@bjtu.edu.cn
* Corresponding author (Ch. Zhang) Email: c.zhang@uni-siegen.de

# 1. Introduction

During the past two decades, the negative index electromagnetic metamaterials have received considerable attention and shown great potential for new and exotic phenomena, because these artificially constructed materials can possess characteristics unlike those of any conventional materials, e.g. negative properties such as negative permittivity and negative permeability, etc. (Pendry et al., 1996, 1999; Smith et al., 2000, 2004; Shelby et al., 2001; Valentine et al., 2008; Xu et al., 2013) These negative properties enable some promising applications such as negative refraction (Shelby et al., 2001; Smith et al., 2004; Hoffman et al., 2007; Valentine et al., 2008; Xu et al., 2013), superlensing (Fang et al., 2005; Smolyaninov et al., 2007; Liu et al., 2007), invisibility cloaking (Lai et al., 2009), subwavelength optical waveguides (Alù and Engheta, 2004), enhanced sensing (Jakšić et al., 2010; Chen et al., 2012) and nonlinear optical generation (Lan et al., 2015), etc. Analogously, inspired by the surprising manipulation of electromagnetic waves, much effort has also been put into the negative index acoustic and elastic metamaterials (EMMs) for various applications. Certainly, their emergence has significantly broadened the full control of acoustic and elastic waves. Due to the resonance-induced effective materials characteristics, i.e., negative mass density (Ding et al., 2007; Lai et al., 2011; Liu et al., 2011a, 2011b; Wu et al., 2011; Yang et al., 2013; Zhu et al., 2014; Liu et al., 2015; Oh et al., 2016a, 2016b; Yang et al., 2016), negative bulk modulus (Ding et al., 2007; Lai et al., 2011; Yang et al., 2013; Zhu et al., 2012, 2014;), and negative shear modulus (Lai et al., 2011; Wu et al., 2011), the acoustic and elastic metamaterials with various physical schemes provided the novel phenomena and applications including negative refraction (Liu et al. 2011a; Wu et al., 2011; Zhu et al., 2014), subwavelength imaging (Kaina et al. 2015), near-filed amplification (Park et al., 2011), and vibration attenuation (Liu et al., 2011b etc.). Generally, both isotropic and anisotropic metamaterials can give rise to the unusual constitutive properties. When the anisotropy becomes strong, a hyperbolic metamaterial (Christensen and García de Abajo, 2012; Poddubny et al., 2013; Shen et al., 2015) is obtained. In the context of metamaterials, structures with the high (Lai et al., 2011; Wu et al., 2011) or reduced symmetry (Liu et al., 2011a, 2011b; Zhu et al., 2014;) were proposed to achieve different monopolar, dipolar, quadrupolar and rotational responses. For the double negativity, it is possible to combine two structures supporting two resonances or construct a structure displaying overlapping responses simultaneously. Meanwhile, this presents new challenges for manufacturing metamaterials with low loss and high accuracy. To make metamaterials more practical, some researchers reported structures with tunable and reconfigurable negative properties (Baz et al., 2009; Kasirga et al., 2009; Ma et al., 2014).

Unlike the electromagnetic and acoustic metamaterials, however, the increased number of material parameters for EMMs will result in more complex wave modes, especially when the double-negative index is considered. Certainly, more complexity also means more possibilities to realize fantastic phenomena, e.g., wave mode conversion from the longitudinal (transverse) wave to transverse (longitudinal) wave at the interface between metamaterial and a natural solid (Wu et al., 2011; Zhu et al., 2014). The three-phase fluid-solid EMMs can produce the simultaneous negative mass density and negative shear modulus (Wu et al., 2011). The four-phase (Lai et al., 2011), three-phase (Liu et al. 2011a) or single-phase (Zhu et al., 2014) solid EMM was also demonstrated to possess a pass band with the negative mass density and negative bulk modulus simultaneously. Obviously, owing to the potential for new functional elastic devices, the double-negative elastic metamaterials will keep attracting more and more researchers.

However, no systematic design strategy has been reported for the double-negative EMMs in the frequency range of interest yet. In other words, no one has considered the inverse problem of this intriguing material. Actually, limited by the symmetries and artificially designed geometries, the existing structures have the relatively narrow double-negative frequency ranges (Lai et al., 2011; Wu et al., 2011; Liu et al. 2011a; Zhu et al., 2014). Moreover, further resonant mechanisms for the double-negative index are needed to be explored to provide more possibilities for designing artificial metamaterials with novel physical properties. On the other hand, most reported double-negative EMMs only have the isotropic (Wu et al., 2011; Lai et al., 2011) or weakly anisotropic (Liu et al. 2011a; Zhu et al., 2014) dispersions, which limits the applicability of double-negative metamaterials. Therefore, it is necessary to further increase anisotropy extent of EMMs. In the last decade, topology optimization as a mathematical approach has been used to successfully discover and propose many outstanding photonic/phononic crystals and devices (Sigmund and Hougaard, 2008; Yamada et al. 2013; Men et al., 2011, 2014; Meng et al. 2015; Sigmund and Jensen, 2003; Bilal et al., 2011; Vatanable et al., 2014; Dong et al., 2014a, 2014b, 2014c; Liu et al., 2016; Yang et al., 2016; Frandsen et al., 2014; Piggott et al., 2015; Isakari et al. 2016), showing the universal design scheme and guiding the corresponding fabrication, simultaneously. Recently, a topology-optimized acoustic structure for negative refractive waves with the desired angle and the corresponding experimental validations have been reported (Christiansen and Sigmund, 2016a, 2016b). Similarly, a new topology optimization algorithm based the bidirectional evolutionary structural optimization has been proposed to create novel photonic crystals with all-angle negative refraction (Meng et al., 2016). At the same time, a level set-based topology optimization method for the design of an optical hyperlens is presented (Otomori et al., 2016). Inspired by these remarkable studies, topology optimization is expected to reveal the beneficial geometries of the negative index EMMs, thus making metamaterials improvable and more practical.

In this paper, for the first time, we perform topology optimization of two-dimensional (2D) anisotropic EMMs with the simultaneous negative mass density and negative longitudinal modulus at the prescribed frequency range. On account of the resonance mechanisms of double-negativity, the optimization objective function is proposed based on the effective medium theory. We successfully design the optimized holey structures made of a single-phase solid material exhibiting the broadband double-negative index under different wavelengths and analyze the mechanism of double-negativity based on the multipolar resonances. The negative refraction and superlensing with breaking diffraction limit at the subwavelength scale are numerically demonstrated. Remarkably, we firstly discover the negative refraction of a transverse wave in the single-phase double-negative EMMs. Meanwhile, most optimized EMMs can serve as zero-index metamaterials (ZIMs) exhibiting cloaking effects for the longitudinal or transverse wave at a certain frequency. Finally, we apply the special constraints in optimization and design a super-anisotropic EMM which yields the double-negative dispersion along one direction and hyperbolic dispersion in the orthotropic direction.

This paper is organized as follows: Section 2 describes the effective medium approach and effective material parameter calculation. Section 3 presents the formulation and procedure of the topology optimization. Then, the results, mechanism analysis and emerging applications of the optimized anisotropic EMMs with broadband double-negative index and super-anisotropic EMM with hyperbolic dispersion are presented in Secs. 4 and 5, respectively. Finally, the conclusions are drawn in Sec. 6.

## 2. Effective medium approach and effective material parameters

Metamaterials are artificial composite media engineered to have properties beyond the nature at a scale much smaller than the wavelength. Figure 1 displays a schematic illustration of a metamaterial composed of infinite periodic microstructures. In the absence of the body forces, the 2D linear elastic harmonic wave equation for a purely elastic medium is given by

$$\nabla \cdot \{[\lambda(\mathbf{r}) + 2\mu(\mathbf{r})](\nabla \cdot \mathbf{u})\} - \nabla \times [\mu(\mathbf{r})\nabla \times \mathbf{u}] + \rho \omega^2 \mathbf{u} = 0, \quad (1)$$

where $\nabla$ is the 2D gradient operator; $\lambda$ and $\mu$ are the position-dependent Lamé constants; $\mathbf{r}=(x, y)$ signifies the position vector; $\mathbf{u}=(u_x, u_y)$ represents the displacement vector; $\rho$ denotes the mass density; and $\omega$ is the circular frequency. The elastic modes in this 2D system can be decoupled into the scalar and vector parts. Their corresponding modes are called the out-of-plane (along the $z$-axis) and in-plane wave modes (in the $xy$-plane). Here, we only study the in-plane wave mode. The relevant decoupled wave equations from Eq. (1) can be expressed as

$$\begin{cases} \rho(\mathbf{r})\omega^2 u_x + \dfrac{\partial}{\partial x}\left[\lambda(\mathbf{r})\left(\dfrac{\partial u_x}{\partial x} + \dfrac{\partial u_y}{\partial y}\right)\right] + \dfrac{\partial}{\partial x}\left[\mu(\mathbf{r})\left(\dfrac{\partial u_x}{\partial x} + \dfrac{\partial u_x}{\partial x}\right)\right] + \dfrac{\partial}{\partial y}\left[\mu(\mathbf{r})\left(\dfrac{\partial u_x}{\partial y} + \dfrac{\partial u_y}{\partial x}\right)\right] = 0 \\ \rho(\mathbf{r})\omega^2 u_y + \dfrac{\partial}{\partial y}\left[\lambda(\mathbf{r})\left(\dfrac{\partial u_x}{\partial x} + \dfrac{\partial u_y}{\partial y}\right)\right] + \dfrac{\partial}{\partial x}\left[\mu(\mathbf{r})\left(\dfrac{\partial u_y}{\partial x} + \dfrac{\partial u_x}{\partial y}\right)\right] + \dfrac{\partial}{\partial y}\left[\mu(\mathbf{r})\left(\dfrac{\partial u_y}{\partial y} + \dfrac{\partial u_y}{\partial y}\right)\right] = 0 \end{cases}. \quad (2)$$

According to the Bloch's theorem (Brillouin 1953), the elastic wave propagating in a periodic structure has the form of $\mathbf{u}(\mathbf{r}) = e^{i(\mathbf{k}\cdot\mathbf{r})}\mathbf{u}_\mathbf{k}(\mathbf{r})$, where $\mathbf{u}_\mathbf{k}(\mathbf{r})$ is a periodic function of the spatial position vector $\mathbf{r}$ with the same periodicity as the structure, and $\mathbf{k}=(k_x, k_y)$ is the Bloch wave vector. For some complex geometries, it is useful to employ the finite element method (FEM) to solve the wave equation incorporating the Bloch condition. The generalized eigenvalue equation in a discrete form can be written as

$$\left[\mathbf{K}(\mathbf{k}) - \omega^2 \mathbf{M}\right]\mathbf{U} = 0, \quad (3)$$

where $\mathbf{K}$ and $\mathbf{M}$ are the global stiffness and mass matrices, respectively; and $\mathbf{U}$ is the column vector formed by all node displacements. In this study, the general FEM software ABAQUS is adopted to solve the eigenvalue equation (2) (Dong et al. 2014a). In light of incapability for the complex eigenvalue problems in ABAQUS, we should divide Eq. (3) into the real and imaginary parts (Dong et al. 2014a)

$$\left(\begin{bmatrix} \mathbf{K}_R & -\mathbf{K}_I \\ \mathbf{K}_I & \mathbf{K}_R \end{bmatrix} - \omega^2 \begin{bmatrix} \mathbf{M}_R & -\mathbf{M}_I \\ \mathbf{M}_I & \mathbf{M}_R \end{bmatrix}\right)\begin{bmatrix} \mathbf{U}_R \\ \mathbf{U}_I \end{bmatrix} = 0, \quad (4)$$

where the subscripts $R$ and $I$ represent the real and imaginary parts, respectively.

Then, the displacement constraint condition based on the Bloch theorem has to be divided into the real and imaginary parts (Dong et al. 2014a)

$$\begin{bmatrix} \mathbf{U}_R(\mathbf{r}) \\ \mathbf{U}_I(\mathbf{r}) \end{bmatrix} = \begin{bmatrix} \cos(\mathbf{k}\cdot\mathbf{a}) & \sin(\mathbf{k}\cdot\mathbf{a}) \\ -\sin(\mathbf{k}\cdot\mathbf{a}) & \cos(\mathbf{k}\cdot\mathbf{a}) \end{bmatrix}\begin{bmatrix} \mathbf{U}_R(\mathbf{r}+\mathbf{a}) \\ \mathbf{U}_I(\mathbf{r}+\mathbf{a}) \end{bmatrix}, \quad (5)$$

where $a$ is the lattice constant. We can solve Eq. (3) combined with Eq. (5) by using ABAQUS/Standard eigenfrequency solver Lanzcos with $\mathbf{k}$ varying within the first Brillouin zone. As a result, the whole dispersion relation ($\mathbf{k}$-$\omega$) is obtained.

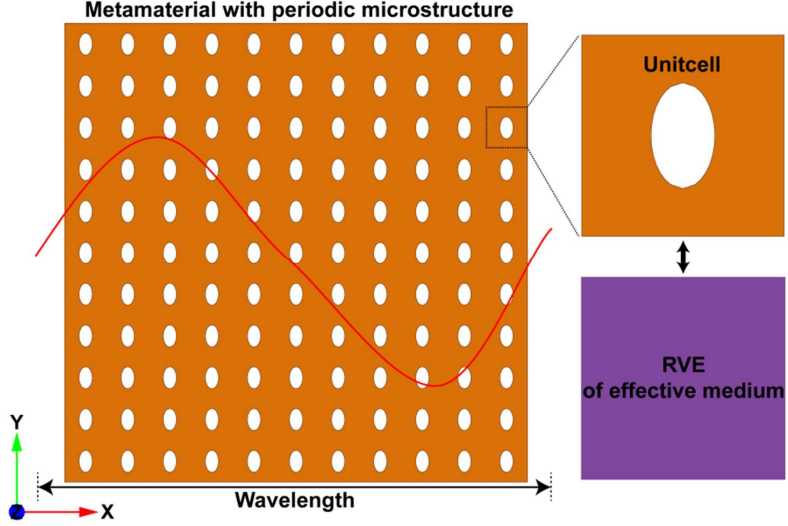

**Fig. 1.** Schematic illustration of a metamaterial with periodic microstructure, the unitcell and the effective medium.

For characterizing the wave propagation in metamaterials, it is very useful to retrieve the effective material parameters by the effective medium theory (Lai et al., 2011; Liu et al., 2011a, 2011b; Zhu et al., 2014). That is, the metamaterial with periodic microstructure can be regarded as a homogeneous effective medium. The periodic unitcell is equivalent to the representative volume element (RVE), see Fig. 1. We can easily understand the behaviors of elastic waves based on the dynamic effective mass density and elastic moduli. Under the long wavelength assumption, the effective medium parameters can be obtained by calculating the boundary states of the metamaterial unitcell (Liu et al., 2011b), i.e., the averaged resultant forces, strains, displacements and accelerations. In other words, the unitcell is taken as the basic RVE which responses to the outside harmonic waves. It is important to note that, unlike the volume average, the effective medium theory based on the surface response of a unitcell is more suitable for describing the strong anisotropic relative motions (Lai et al., 2011). Generally, we apply the suitable displacement field on the boundaries of a unitcell and calculate the effective physical parameters of the whole metamaterial structure. Since the effective parameters mainly depend on the operating frequencies, we do not consider the role of the wave vector for the sake of simplicity. In accordance with the Newton's second law and the energy balance principle, the effective mass density and elastic moduli can be further determined, respectively. Next we describe briefly the calculation procedure of the effective material parameters. For details, we refer to Liu et al. (2011a, b) and Lai et al. (2011).

In view of the strong anisotropy and large search space, we consider the metamaterial unitcell in a square lattice with the orthogonal symmetry for the optimization design in this paper, as shown in Fig. 2(a). In order to evaluate the effective mass density, the nodes of the unitcell boundaries (i.e., L, R, D and U in Fig. 2(a)) are assumed to be subjected to a global displacement $\mathbf{u} = \mathbf{U}^0 e^{i\omega t}$ when the displacement phase difference is ignored (Liu et al., 2011a, 2011b). According to the Newton's second law, the effective mass density tensor is easily computed from (Liu et al., 2011b; Zhu et al., 2012; Oudich et al., 2014)

$$\begin{bmatrix} F_1^* \\ F_2^* \end{bmatrix} = -\omega^2 V \begin{bmatrix} \rho_{11} & 0 \\ 0 & \rho_{22} \end{bmatrix} \begin{bmatrix} U_1^0 \\ U_2^0 \end{bmatrix}, \tag{6}$$

where $F_{i=1,2}^*$ and $U_{i=1,2}^0$ are the induced forces and applied displacement field (Liu et al., 2011b) on the boundaries, respectively; $\omega$ is the angular frequency; $V$ is the volume of the effective medium; and $(\rho_{ij})_{i,j=1,2}$ indicates the effective anisotropic mass density matrix.

For calculation of effective moduli, however, the special displacement fields associated with the eigenstates are needed to be applied to the discrete boundary nodes of the unitcell. Considering different deformations, we can write the global strain field as

$$\mathbf{E}^0 = \begin{bmatrix} \varepsilon_{11}^0 & \varepsilon_{12}^0 \\ \varepsilon_{21}^0 & \varepsilon_{22}^0 \end{bmatrix}, \tag{7}$$

where $(\varepsilon_{ij}^0)_{i,j=1,2}$ represents the applied strain field. Then applying the corresponding displacement field $\mathbf{u} = \mathbf{E}^0 \times \mathbf{x} e^{i\omega t}$ (where $\mathbf{x}$ is the coordinate vector) on the boundary nodes. We can obtain the components of the elastic moduli from the equivalence (Liu et al., 2011b) between the energy of the induced force on the unitcell and the strain energy of the effective medium, i.e.,

$$\sum_{\text{LRDU}} \mathbf{F}^* \times \mathbf{u}^* = C_{ijkl} \varepsilon_{ij}^0 \varepsilon_{kl}^0 V, \tag{8}$$

where the subscript LRDU represents the whole boundary edges (L, R, D and U in Fig. 2(a)); $\mathbf{F}^*$ and $\mathbf{u}^*$ denote the induced force and displacement vectors; $i, j, k, l$=1, 2; and $(\varepsilon_{ij}^0)_{i,j=1,2}$ and $(\varepsilon_{kl}^0)_{k,l=1,2}$ are the applied strain fields (Liu et al., 2011b). For the 2D metamaterial with the orthogonal symmetry shown in Fig. 2(a), the nonzero effective moduli are $C_{1111}=C_{11}$, $C_{2222}=C_{22}$, $C_{1212}=C_{44}$ and $C_{1122}=C_{12}$. For some special cases, the effective bulk and shear moduli, $K_{eff}$ and $\mu_{eff}$ are also useful to analyze the equivalent behaviors (Lai et al., 2011; Liu et al., 2011b). Their corresponding applied strain fields are, respectively,

$$\mathbf{E}^0 = \begin{bmatrix} 1 & 0 \\ 0 & 1 \end{bmatrix} \text{ and } \mathbf{E}^0 = \begin{bmatrix} 1 & 0 \\ 0 & -1 \end{bmatrix}. \tag{9}$$

For simplicity, we define the bulk and shear moduli (Lai et al., 2011; Liu et al., 2011b) respectively as

$$K_{eff} = \frac{C_{11} + 2C_{12} + C_{22}}{4}, \quad \mu_{eff} = \frac{C_{11} - 2C_{12} + C_{22}}{4}, \tag{10}$$

which are different from the traditional definitions in the linear elasticity theory. Therefore, the strain energy $U^*$ for the volume deformation and shear deformation are respectively formulated from Eq. (8) as

$$U^* = 4K_{eff}V, \quad U^* = 4\mu_{eff}V. \tag{11}$$

After obtaining the effective mass density and elastic modulus tensors, we can easily write the corresponding Christoffel equation for the plane harmonic bulk waves in the anisotropic elastic metamaterials as

$$(C_{ijkl}n_k n_l - v^2 \rho_{il})U_l = 0, \tag{12}$$

where $i, j, k, l$=1, 2; $n_k$ and $n_l$ are the propagation vector components; $v$ represents the phase velocity; and $U_l$ denotes the displacement amplitude. The effective phase velocity of the longitudinal wave propagating along $\Gamma X$ direction is given by $v_l = \sqrt{C_{11}/\rho_{11}}$.

Since the effective parameters reveal the resonance of the microstructures, the combinations of negative or positive mass density $\rho_{11}$ and longitudinal modulus $C_{11}$ can give rise to interesting physical properties. For instance, a single negative $\rho_{11}$ or $C_{11}$ in a frequency region will lead to imaginary $v_l$, which implies the existence of a bandgap for the longitudinal wave. The previous studies showed that the negative mass density is usually created by the dipolar resonances (Liu et al., 2011a, 2011b; Lai et al., 2011; Wu et al., 2011; Zhu et al., 2014). However, the monopolar and quadrupolar resonances produce the negative bulk modulus and negative shear modulus, respectively (Liu et al., 2011a; Lai et al., 2011; Wu et al., 2011; Zhu et al., 2014). Of course, the simultaneous positive or negative $\rho_{11}$ and $C_{11}$ mean that the longitudinal wave can propagate through the metamaterial. However, the simultaneous negative $\rho_{11}$ and $C_{11}$ will induce negative effective refractive index for the longitudinal wave, that is, the left-hand property $\mathbf{S}\cdot\mathbf{k}<0$ (where $\mathbf{S}$ and $\mathbf{k}$ represent the Poynting vector and wave vector, respectively) can be achieved for the refractive waves. For anisotropic materials which will be considered in this paper, more material parameters will be involved in the phase velocity, resulting in more complex dispersion relations. Therefore, the anisotropic EMM is expected to reveal new resonances and produce more interesting novel wave behaviors.

## 3. Formulation and procedure of the topology optimization

The present paper will perform topology optimization for a 2D anisotropic EMM based on the effective material parameters. To this end, we need first construct a suitable objective function with proper constraints.

### 3.1. Objective function and constraints

To clearly illustrate the motivation of development of the objective function, we follow the reported works about the physical mechanisms of generation of the negative mass density $\rho_{\text{eff}}$ and negative elastic modulus $C_{\text{eff}}$ (Liu et al., 2011a, 2011b; Lai et al., 2011; Wu et al., 2011; Zhu et al., 2014), and show the evolution of $\rho_{\text{eff}}$ and $C_{\text{eff}}$ from the positive values to negative ones with variation of the EMM unitcell's topology in Figs. 2(b) and 2(c) (the evolution direction is from step 1 to step 3). We first consider the effective mass density $\rho_{\text{eff}}$. At the initial stage, a single-phase material without any hole (i.e., with a zero porosity) has a constant $\rho_{\text{eff}}$, see step 1 in Fig. 2(b). When the porosity increases with the unitcell's topology being modified, appearance of a resonance at a certain frequency $f_{R\rho}$ in the operating frequency region is necessary for generating a negative $\rho_{\text{eff}}$ (Liu et al., 2011a, 2011b; Lai et al., 2011; Wu et al., 2011; Zhu et al., 2014), see step 2 in Fig. 2(b). Different effective resonance directions below and above the resonant frequency $f_{R\rho}$ result in a jump of $\rho_{\text{eff}}$ from a positive infinity to a negative infinity in a very narrow frequency range. Then the operating frequency region is divided into four regions (marked by the four color bars in Fig. 2(b)): (1) the statics limiting region (L) where $\rho_{\text{eff}}$ increases with the frequency and decreases with the porosity increasing; (2) the region below $f_{R\rho}$ ($M_1$) where $\rho_{\text{eff}}$ increases to the positive infinity with the frequency rising; (3) the region above $f_{R\rho}$ ($M_2$) where negative $\rho_{\text{eff}}$ is produced and increases from the negative infinity with the frequency increasing, turning from negative to positive value at the frequency $f_{0\rho}$; and (4) high frequency region (U) where $\rho_{\text{eff}}$ increases continuously. It is seen that we have a negative $\rho_{\text{eff}}$ range ($f_{R\rho}$, $f_{0\rho}$). If the curves of $\rho_{\text{eff}}$ move to left with $f_{R\rho}$ unchanged (possibly due to variation of the unitcell's topology, e.g. from step 2 to step 3), the negative $\rho_{\text{eff}}$ range will becomes wider. If the resonant frequency decreases (e.g. $f'_{R\rho}$ of step 3 in Fig. 2(b)), the negative $\rho_{eff}$ range will be lower. That is to say, if a lower negative $\rho_{eff}$ range is desired, we should alter the unitcell's topology to drive the curve in range L left as well as drive down $f_{R\rho}$ as much as possible.

Besides, we display the generation of a negative effective modulus $C_{eff}$ in Fig. 2(c). At the initial stage (step 1) when a single-phase material is homogeneous with a zero porosity, $C_{eff}$ does not change with the frequency. With the change of the EMM unitcell's topology, the performance of $C_{eff}$ is likely to vary from step 1 to step 2. To produce negative $C_{eff}$, a resonance at a certain frequency $f_{RC}$ should appear in the operating frequency region (Liu et al., 2011a; Lai et al., 2011; Wu et al., 2011; Zhu et al., 2014), see step 2 in Fig. 2(c). We again have four regions (L, $M_1$, $M_2$ and U) as in Fig. 2(b). Unlike $\rho_{eff}$ in Fig. 2(b), $C_{eff}$ in region L decreases with the frequency increasing and reaches zero at $f_{0C}$. In region $M_1$, $C_{eff}$ decreases to a negative infinity as the frequency rises, and then changes abruptly to a positive infinity across $f_{RC}$. Thus, ($f_{R\rho}$, $f_{0\rho}$) is a negative $C_{eff}$ range. With the frequency increasing in region $M_2$, $C_{eff}$ decreases from the positive infinity and then becomes negative in region U. To broaden the negative $C_{eff}$ range, it is necessary to revise the unitcell's topology to move $C_{eff}$ curves left (e.g. from step 2 to step 3). Meanwhile, the resonant frequency will be driven downward (e.g. $f'_{RC}$ in step 3 in Fig. 2(c)).

Therefore, a basic condition for generating negative $\rho_{eff}$ or $C_{eff}$ is to excite a resonance and try to make the positive or negative infinities to appear at low frequencies. And a helpful way to expand the negative ranges at lower frequency regions for both $\rho_{eff}$ and $C_{eff}$ (e.g. expand ($f_{R\rho}$, $f_{0\rho}$) to ($f'_{R\rho}$, $f'_{0\rho}$), or ($f_{0C}$, $f_{RC}$) to ($f'_{0C}$, $f'_{RC}$)) is to design a suitable EMM unitcell's topology to move the curves of $\rho_{eff}$ and $C_{eff}$ left and push down the resonant frequency $f_R$. Particularly, if a wide low frequency range of double negative $\rho_{eff}$ and $C_{eff}$ is concerned, we should have a large overlapping region between the negative $\rho_{eff}$ region ($f'_{R\rho}$, $f'_{0\rho}$) and negative $C_{eff}$ region ($f'_{0C}$, $f'_{RC}$) (see step 3 in Fig. 2(b) and 2(c)). That is to say, the existence condition for double-negativity is $\min\{f'_{0\rho}, f'_{RC}\} > \max\{f'_{0C}, f'_{R\rho}\}$. Consequently, in implementation of topology optimization, our concern may be focused on the frequency region below $f_{RC}$.

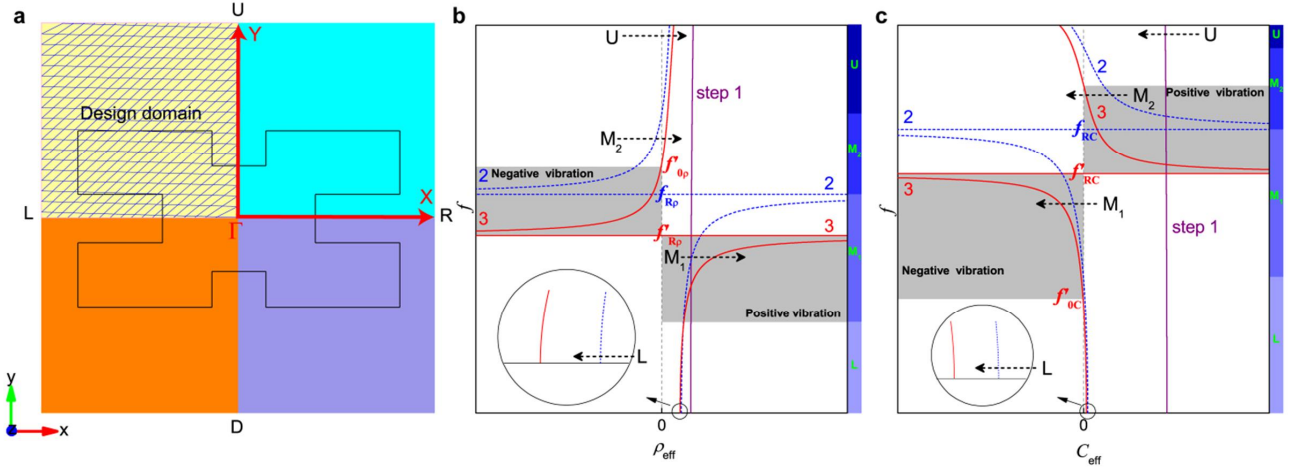

**Fig. 2. Optimization model in the square lattice and illustration for the negative index formation. (a)** Unitcell with the orthogonal symmetry and the design domain in optimization; **(b)** evolution of the effective mass density; and **(c)** evolution of the elastic modulus. The principle directions (ΓX and ΓY) of the first Brillouin zone are shown as well. The inner geometry in (a) shows an orthogonal example. The numbers of 1, 2 and 3 in (b) and (c) represent the evolution order as the possible EMM topology changes. The color bars (L, $M_1$, $M_2$ and U) in (b) and (c) represent the divided frequency regions for step 2.

For the present optimization problem, we have two objective properties: negative effective mass density $\rho_{eff}$ and negative effective elastic modulus $C_{eff}$ which appear in the same frequency region. However, the evolution trends of the unitcell's topology for these two properties may be inconsistent or even conflicting. Fortunately, topology optimization methods can effectively solve the problem with inconsistent or even conflicting objectives and find out the possible solution which balances two factors appropriately. For this end, we should first propose a proper objective function.

In this paper, we are pursuing the double-negativity for longitudinal wave propagation along ΓX direction. Therefore we will apply topology optimization method to design a 2D holey structure which exhibits both negative effective mass density $\rho_{11}$ and negative effective longitudinal modulus $C_{11}$ simultaneously in a wide enough frequency region within a prescribed operating frequency range, i.e. the target frequency range, ($f_{min}$, $f_{max}$), where it is understood that the wavelength is far longer than the lattice constant. To construct a proper object function, we select some sampling frequencies distributed uniformly in ($f_{min}$, $f_{max}$). It is noticed from Figs. 2(b) and 2(c) that the minimal positive value of $\rho_{eff}$ appears at zero-limiting frequency or near $f'_{0\rho}$ and the maximal positive value appears below $f'_{R\rho}$. Below $f'_{RC}$, the maximal positive value of $C_{eff}$ occurs at the zero-limiting frequency, and the minimal positive value occurs near $f'_{0C}$. Based on the above analysis, we can find that a possible way to realize the evolution from a positive index to a negative one (e.g. from step 2 to step 3) is to enlarge the gap between its maximal and minimal positive values (i.e. to increase the ratio between maximal value to minimal value for all sampling frequencies). Inspired by all these considerations, we suggest the following objective function for the present optimization problem:

$$For: f_{min} \to f_{max}, \tag{13}$$

$$Maximize: D = ND + \min \left\{ -\alpha \cdot \frac{\min_{\forall m(m \in (1,2,\ldots,M))} \left( \rho_{11}^+(m) \right)}{\max_{\forall m(m \in (1,2,\ldots,M))} \left( \rho_{11}^+(m) \right)}, -\beta \cdot \frac{\min_{\forall n(n \in (1,2,\ldots,M))} \left( C_{11}^+(n) \right)}{\max_{\forall n(n \in (1,2,\ldots,M))} \left( C_{11}^+(n) \right)} \right\}, \tag{14}$$

where $M$ is the number of the sampling frequencies within the target operating frequency range ($f_{min}$, $f_{max}$); $D$ is the proposed objective function value; $ND$ declares the number of the sampling frequencies where simultaneous negative $\rho_{11}$ and $C_{11}$ are generated; $\alpha$ and $\beta$ ($\alpha+\beta=1$) are the weight coefficients of the functions for $\rho_{11}$ and $C_{11}$, respectively (selection of values of $\alpha$ and $\beta$ will be discussed later); $\rho_{11}^+$ and $C_{11}^+$ indicate the special arrays composed of the positive $\rho_{11}$ and positive $C_{11}$, respectively, among the total sampling frequencies; $m$ ($m \leq M$) is the serial number of the sampling frequencies where the positive $\rho_{11}$ remains; and $n$ ($n \leq M$) is the serial number of the sampling frequencies where the positive $C_{11}$ exhibits. If $f_{min}=0$, the computation cost for optimization will be too high because ABAQUS/Standard solver cannot directly solve the steady-state dynamic response at 0 Hz. So we select the value of $f_{min}$ as 0.5 Hz instead of 0 Hz to reduce the computational cost.

It is noticed that we use $-\min()/\max()$ instead of $\max()/\min()$ in Eq. (14) to avoid possible ill-condition in calculation because the minimal positive values for $\rho_{11}^+$ and $C_{11}^+$ are possible to be extremely small. The number of the sampling frequencies with the double-negativity (*ND*) appearing in Eq. (14) is to ensure the double-negative metamaterial to be completely superior to others. With the emerging of negative values, both *ND* and *D* will be improved evidently, guiding the optimization algorithm to explore more negative values. Here, we specially stress the importance of the weight coefficients $\alpha$ and $\beta$, the sum of which is normalized to 1. Since the optimization involves two objectives, how to explore their trade-off relation becomes the key mechanism to find the resonances for $\rho_{11}$ and $C_{11}$ simultaneously. In optimization, $\alpha$ and $\beta$ can control the impact degrees of $\rho_{11}$ and $C_{11}$ on the evolution, respectively. The larger $\alpha$ (smaller $\beta$) or larger $\beta$ (smaller $\alpha$) will force the optimization algorithm to give priority to the variation of $\rho_{11}$ or $C_{11}$ respectively. Certainly, if $\alpha$ equals $\beta$, the variation for $\rho_{11}$ and $C_{11}$ are treated equally. Once the negative $\rho_{11}$ is produced, $C_{11}$ will be the main concern objective to make $C_{11}$ negative as well, and vice versa. Through the repeated balances, *D* gradually increases from a negative value to a large positive one.

For some physical reasons, we need to introduce the following three special constraints in optimization.

(1) If the rotation of the unitcell's boundary is very strong, the classic linear effective medium theory used in this paper cannot characterize the motion of the unitcell accurately. In other words, the effective mass density tensor can no longer be diagonal. Therefore, we should apply a constraint to make sure that the translational motion dominates the boundary's actions without obvious local rotation. At the same time, this also leads to the obvious polarization of the longitudinal wave along $\Gamma X$ direction. This constraint is stated as

$$\textbf{\textit{Subject to}}: mF_y = \max_{\forall i(i=1,2,\ldots,M)} \left\{ \left( \frac{\sum_{LR}|F_2|}{\sum_{LR}|F_1|} \right) \right\} \leq \delta_F, \tag{15}$$

where $mF_y$ describes the vibration along *y*-direction; *i* (*i*=1, 2,..., *M*) is the serial number of the sampling frequency; subscript LR represents the nodes of boundaries L and R in Fig. 2(a); $F_1$ and $F_2$ are, respectively, the magnitudes of the reaction forces along *x*- and *y*-directions when calculating $\rho_{11}$; $\delta_F$ is a vibration control factor which is positive. To avoid unexpected stop of the optimization evolution, we suggest that the positive $\delta_F$ is better to takes a value smaller than 0.25.

(2) We have to use another constraint to control the coupling modulus $C_{12}$ between two principle directions. Along with the decreasing of $C_{11}$, the value of $C_{12}/|C_{11}|$ will change non-monotonously. However, if $C_{12}/|C_{11}|$ becomes very small, it is easier to bring about the vibration along the $\Gamma Y$ direction when the vibration along $\Gamma X$ direction is excited. That is, the coupling is very strong, which will lead to complex wave motions. Because our goal is to make the longitudinal wave propagate along $\Gamma X$ direction, more strong vibrations along the $\Gamma Y$ direction will result in more complex propagating features: both the longitudinal and transverse vibrations along the $\Gamma Y$ direction may be induced by the propagating waves along $\Gamma X$ direction. Especially, the transverse wave mode is possible to be excited and coupled with the longitudinal wave mode when $C_{12}$ turns negative (we will further demonstrate this phenomena in Sec. 4.2.). Hence, compared with the absolute value of $C_{11}$, it is better to guarantee that $C_{12}$ is not very small and always positive within the target operating frequency range. This constraint takes the form of

$$\textbf{\textit{Subject to}}: mC_{12} = \min_{\forall i(i=1,2,\ldots,M)} \left[ \left( \frac{C_{12}}{|C_{11}|} \right) \right] \geq \delta_C, \tag{16}$$

where $mC_{12}$ describes the difference level between $C_{12}$ and $|C_{11}|$; $\delta_C$ signifies an important positive value that controls the coupling degree. Several selections of $\delta_C$ will be showed to explain its effect in Sec. 4.3.1. In general, we suggest that $\delta_C$ should be larger than 0.1.

(3) We have to control the minimal size of the optimized structure aiming at overcoming the mesh-dependence problem (Sigmund and Petersson, 1998) in optimization and realize the manufacturable structure. Although the narrow connections may widen the double-negative range, we still propose that the minimal width should be larger than a prescribed limited width. More details on this constraint can be found in our previous works (Dong et al., 2014b, 2014c). The corresponding geometrical constraint is

$$\textbf{\textit{Subject to}}: \min_{\phi}(w) \geq w^*, \tag{17}$$

where $\phi$ is the topological distribution within the unitcell; $w$ is the array composed of the width of every solid connection; and $w^*$ expresses the minimal size control parameter.

In summary, the design of an EMM supporting simultaneous negative $\rho_{11}$ and $C_{11}$ in a frequency range as wide as possible is formulated as a topology optimization problem with the objective function in Eqs. (13) and (14) with constraints in Eqs. (15)-(17).

### 3.2 Procedure of topology optimization

To solve the optimization problem in Eqs. (13)-(17), we adopt the single-objective genetic algorithm (GA) (Holland, 1975; Dong et al., 2014a) as the inverse method and the finite element software ABAQUS/Standard solver for the effective parameters calculation. GA, one of the most popular evolutionary algorithms, generates solutions to optimization problems using techniques inspired by natural evolution. Each design is characterized by a chromosome which contains a number of genes describing the whole structure geometry. Unlike the gradient-based optimization methods (Sigmund and Hougaard, 2008; Men et al., 2011, 2014), the fitness is the unique driving force of GA because the basic strategy of GA is the survival of the fittest. For all the optimization problems involving different backgrounds, the unified critical measure is how to accurately evaluate the fitness for an arbitrary structure. In this paper, the corresponding binary GA procedure about the metamaterial design is shown in Fig. 3. In order to get the smooth edges, we initialize our procedure with a coarse grid $N_1 \times N_1$. After $EN_1$ generations, the optimized solution is obtained in the first round of optimization. Then, the optimized structure is mapped into a fine grid $N_2 \times N_2$ ($N_2 > N_1$). Using the refined structure as a seed structure (Dong et al., 2014a), we get the final optimized structure after $EN_2$ generations. The values of $EN_1$ and $EN_2$ are problem-dependent. In this paper, our numerical tests show that $EN_1 = EN_2 = 1000$ can ensure good convergence. After the whole optimization terminates, the final optimized EMM with double-negative index is obtained. The details of the procedure are as follows:

(i) Start with a coarse mesh $N_1 \times N_1$ and a random initial population with $N_p$ individuals, in which every involved metamaterial represents an individual, i.e., the chromosome in the genetic system. The binary coding is used to code the solid (1) and vacuum (0) phases.

(ii) For each individual, apply the "abuttal entropy filter" (Dong et al., 2014c) to filter the geometry at some extent, i.e., some isolated voids (0) are filled up and some isolated solid elements (1) are removed. For any solid element (1), if the number of its 4-neighborhood elements is no more than one, it is changed into a vacuum element (0). Similarly, for any vacuum element (0), if the number of its 4-neighborhood elements is no more than one, it is transferred to a solid element (1). Every individual can be revised to a more simple and clear topology through this approach.

(iii) For each metamaterial, compute the fitness based on the objective function in Eq. (14) and the other parameters in constraints of Eqs. (15)-(17), and form the constraint optimization version. Although we only focus the double-negativity ($\rho_{11}$ and $C_{11}$) in this paper, the user can select different effective parameters ($\rho_{11}$, $\rho_{22}$, $C_{11}$, $C_{22}$, $C_{12}$, $K$ and $\mu$) according to the corresponding metamaterial design demands. Here, we present our fitness expression used in the optimization algorithm as

$$fitness(f) = D(f) - N_{uc}(f) \times 2(M-1) - N_{nw}(f) \times 2(M-1), \tag{18}$$

where $N_{uc}$ is the number of the "disconnected solid areas"; and $N_{nw}$ is the number of the elements which do not satisfy the constraint condition in Eq. (17). In fact, unlike the solid-solid structure, it is very common in optimization that the porous structure is not connected and has many isolated solid regions. Since the maximum of $D$ is larger than ($M$-1) and smaller than $M$, we have to make sure that any solution extremely violating the basic structural constraints should be punished during optimization, no matter how large the objective function value it has is. Therefore, two penalty coefficients $N_{uc}$ and $N_{nw}$ are employed to achieve the optimized connected structure whose minimal size of every part retains a realistic level. In spite of transforming the constraint in Eq. (17) to the penalty coefficient, we also need to take the constraints in Eqs. (15) and (16) into the final fitness evaluation. For every individual in optimization, firstly, we have to judge whether it is a feasible solution or not after obtaining the fitness expression in Eq. (18), that is, check whether it satisfies the constraints in Eqs. (15) and (16) at the same time. There are three cases for an individual:

(a) It is a feasible solution. Then, its final fitness $Ffitness$ equals to the fitness shown in Eq. (18), i.e.,

$$Ffitness = fitness. \tag{19}$$

(b) It is an infeasible solution and no feasible solution exists for the current population. Then, the final fitness of each individual is defined as

$$Ffitness = \begin{cases} \min(fitness) - |mF_y - \delta_F|, & \text{if } mF_y > \delta_F \\ \min(fitness) - |mC_{12} - \delta_C|, & \text{if } mC_{12} < \delta_C \\ \min(fitness) - |mF - \delta_F| - |mC_{12} - \delta_C|, & \text{if } mF_y > \delta_F \text{ and } mC_{12} < \delta_C \end{cases}. \tag{20}$$

(c) It is an infeasible solution and at least one feasible solution exists for the current population. Then, the final fitness of each individual is given by

$$Ffitness = \begin{cases} mfea - |mF_y - \delta_F|, & \text{if } mF_y > \delta_F \\ mfea - |mC_{12} - \delta_C|, & \text{if } mC_{12} < \delta_C \\ mfea - |mF_y - \delta_F| - |mC_{12} - \delta_C|, & \text{if } mF_y > \delta_F \text{ and } mC_{12} < \delta_C \end{cases}, \tag{21}$$

where $mfea$ is the minimum fitness of the feasible solutions in the current population.

(iv) For the current population, using the final fitness $Ffitness$ calculated in step (iii), perform the tournament selection (reproduction) with the size of the competition group in the tournament selection $N_{ts}$, crossover with the crossover probability $P_c$ and mutation with the mutate probability $P_m$ to produce the offspring population.

(v) Use the elitism strategy to accelerate optimization (Dong et al., 2014a). That is, preserve the best individual in the current generation. Then, this elitist will displace the worst individual of the next generation.

(vi) If a fixed number of generations are finished, then stop and return an optimized structure, otherwise, go to step (ii).

(vii) Map the optimized structure into the finer mesh $N_2 \times N_2$ and run a new round of optimization. The optimized structure in the coarse step is used as the seed structure (Dong et al., 2014a) in this fine step.

More details of the optimization procedure can be found in our previous works (Dong et al., 2014a, 2014c).

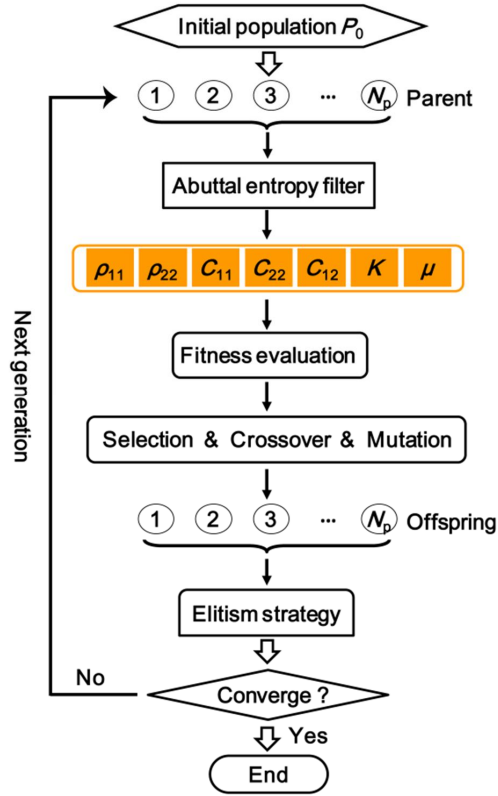

**Fig. 3.** Flow chart of the optimization procedure for metamaterial design. According to the optimization objectives, the corresponding effective parameters ($\rho_{11}$, $\rho_{22}$, $C_{11}$, $C_{22}$, $C_{12}$, $K$ and $\mu$) are considered for the fitness evaluation.

## 4. Optimized anisotropic EMM with broadband double-negative index

In this section, we present optimized results for the square-latticed single-phase metamaterials. The lattice constant $a$=0.03 m is assumed. The material parameters of the stainless steel are the mass density $\rho$=7850 kg m$^{-3}$, the Young's modulus $E$=200 GPa and the poisson's ratio $\upsilon$=0.3 (Zhu et al., 2014). The algorithm parameters of GA are: the population size $N_p$=30, the crossover probability $P_c$=0.9, the mutate probability $P_m$=0.02 for the coarse mesh ($EN_1$=1000) and $P_m$=0.005 for the fine mesh ($EN_2$=1000), and the size of the competition group in the tournament selection $N_{ts}$=18. The coarse and fine meshes are $N_1$=30 and $N_2$=60, respectively. A too small or big $\delta_F$ will result in obvious decrease or increase of the search space. Our numerical tests show that it is effective to use the factor $\delta_F=0.25$ and the number of the sampling frequencies $M=11$ for all optimization cases in this paper. Of course, one can adopt more sampling frequencies. But the optimization results are almost the same. The minimal size control parameter $w^*$ is selected as $a/30$ in this paper. All numerical computations are accomplished on a Linux cluster with 16 cores of Intel Xeon E5-2660 at 2.20 GHz. The results with various optimization parameters, i.e., $f_{max}$, $\delta_C$ and the combination of $\alpha$ and $\beta$, are presented. The double-negative properties of the optimized EMMs are demonstrated by the dispersion curves, vibration modes, effective parameters, group velocities, transmission, negative refraction and imaging. The negative refraction and imaging are simulated by COMSOL Multiphysics 4.4. The other physical calculations are executed by ABAQUS/Standard solver.

## 4.1. Optimized structure with double-negative bands

First, we present the optimized results for the target operating frequency range of $(f_{min}, f_{max})=(0.5$ Hz-19.5 kHz$)$ with the optimization parameters $\alpha=0.5$, $\beta=0.5$ and $\delta_C=0.3$. The longitudinal wavelength of the upper operating frequency is 0.3 m, which is 10 times larger than the lattice constant $a$ (=0.03 m). Figures 4(a) and 4(b) show the optimized 3×3 lattice structure and the optimization evolutions in the coarse and fine steps, respectively. From Fig. 4(a) we can observe that the unitcell structure (shown by the dashed square) is composed of four small solid blocks in the corners and four big lumps in the center. These eight solid parts are connected by eight narrow connections. The four inner solid lumps show obvious anisotropy along $x$- and $y$-directions. And the lattice structure can also be regarded as a periodic array of one rectangular solid block connected with four big solid lumps. These topology features show the typical characteristic of local resonance, opening bandgaps easily. The multiple solid blocks provide much possibility for different resonances. The curve in Fig. 4(b) represents the evolution history of the best structure in the current generation as the generation number increases. Initially, we use the pure solid structure as the seed structure (S1). The change from structure S1 to S2 demonstrates that removing the boundary solids is essential to excite the positive vibration of $\rho_{11}$ and negative vibration of $C_{11}$ simultaneously. So, thinner boundaries will bring out the better fitness, see structures S3 and S4. In particular, we note an obvious jump of evolution curve from structure S4 to S5. This is because two sampling frequencies with the double-negativity is produced in structure S5, i.e., $ND=2$. The only difference between structures S4 and S5 is that the four solid corner blocks become much smaller. This again indicates the important role of the boundaries on the negative property. In the coarse step, the evolution begins to converge at the 752th generation. In the fine step, the structures S6 and S7 have finer edges. The new jump of the evolution curve at the 1010th generation is attributed to a new sampling frequency with double-negativity. As a result, the whole evolution converges at the 1297th generation. The evolution results shown in Fig. 4(b) demonstrate the effectiveness and good convergence of the optimization algorithm. In addition, the results also prove the rationality of the proposed objective function and constraints in Eqs. (13)-(17).

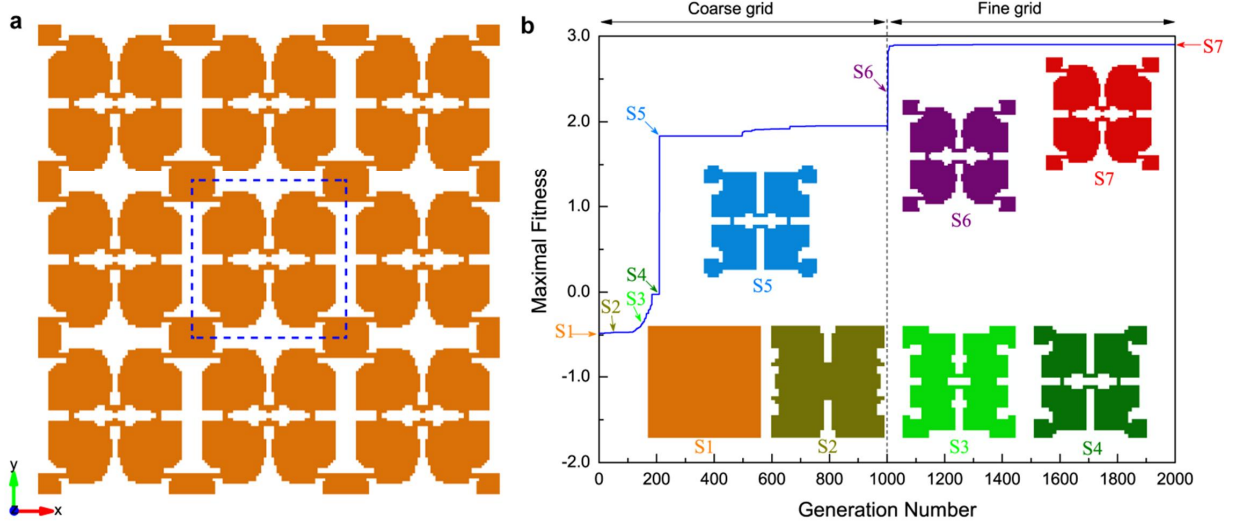

**Fig. 4.** Optimized EMM and the optimization evolution history ($f_{min}$=0.5 Hz, $f_{max}$=19.5 kHz, $\alpha$=0.5, $\beta$=0.5 and $\delta_C$=0.3). **(a)** Final 3×3 lattice structure (the unitcell is enclosed by the dashed line) and **(b)** the unitcell evolution with the increase of the generation number. Seven representative structures are presented as well.

For the optimized EMM in Fig. 4(a), we display the dispersion diagram and effective material parameters in Fig. 5. The structure possesses a pass band (14.52-20.57 kHz) with the negative slope. It is very interesting to observe that the EMM has three negative bands (14.52-20.57 kHz, 24.21-28.34 kHz and 36.48-39.31 kHz). We name these bands L, qT, and T, respectively. To illustrate general idea about the vibration motions, we define a normalized quantity $q_x$ associated with the displacement component $u_x$ along $x$-direction as

$$q_x = \frac{\left|\sum_{unitcell} u_x\right|}{\sqrt{\left(\sum_{unitcell} u_x\right)^2 + \left(\sum_{unitcell} u_y\right)^2}}, \tag{22}$$

where $u_y$ is the displacement component along $y$-direction; and the summation is taken over the whole unitcell nodes. The results of $q_x$ are shown by the colors on bands. Clearly, the first negative band (L) has only the effective vibration motions along $x$-direction. In other words, the global motions within the unitcell can be regarded as the longitudinal wave motions. The second (qT) and third (T) bands have the similar vibration polarizations along $y$-direction. Their effective motions can be regarded as the transverse wave motions.

For verifying the results in Fig. 5(a), we plot the effective material parameters of $\rho_{11}$, $\rho_{22}$ in Fig. 5(b) and $C_{11}$, $C_{22}$, $C_{12}$, $K$, $\mu$ in Fig. 5(c). We find obvious anisotropy of the mass density. The negative $\rho_{11}$ occurs in the frequency range of 14.52-20.57 kHz which is precisely the same as the range of the first negative band (L) as shown in Fig. 5(a). However, $\rho_{22}$ become negative from 14.8 to 40.38 kHz. Figure 5(c) also shows the stronger anisotropy of the modulus. The longitudinal modulus $C_{11}$ firstly changes into negative values in the range of 15.09-33.94 kHz. The upper boundary of this range is almost the same as the lower edge of the sixth band in Fig. 5(a). This coincidence means that this nearly flat band is due to the transformation from negative to positive values in a narrow frequency range. However, it is noted that, the lower boundary (15.09 kHz) is different from the lower edge (14.52 kHz) of the negative $\rho_{11}$ range. This means that the boundary vibration in the frequency range from 14.52 to 15.09 kHz is not the purely longitudinal wave motion although the whole effective vibration is along $x$-direction, that is to say, the eigenstate within the region of 14.52-15.09 kHz cannot be characterized by $\rho_{11}$ and $C_{11}$ only. But, compared with the width (6 kHz) of the negative band, this region with the width of 0.57 kHz is very narrow, and thus can be ignored. Consequently, we obtain a double-negative frequency range of 15.09-20.57 kHz. Because of the complex vibration and strong anisotropy of EMM, we use the overall effective shear modulus $\mu$ to describe the effective shear wave motions. In particular, the effective moduli $C_{12}$ and $\mu$ have the similar variation versus the frequency. $\mu$ and $C_{12}$ turn to be negative from 21.35 kHz and 25.88 kHz, respectively. Therefore, the transverse wave with imaginary $v_t = \sqrt{\mu/\rho_{22}}$ cannot propagate in the double-negative range of 15.09-20.57 kHz. This proves that the first negative band (L) in Fig. 5(a) can only support the longitudinal wave. For the second negative band (qT) in Fig. 5(a), $C_{11}, C_{12}, \rho_{22}$ and $\mu$ are negative, but $\rho_{11}$ is positive and thus $v_l = \sqrt{C_{11}/r_{11}}$ is imaginary. Therefore this band can only sustain the transverse wave. Moreover, we notice the simultaneous negative $\rho_{22}$ and $\mu$ in the region of the third negative band (36.48-39.31 kHz). In the same region, the EMM has positive $\rho_{11}$ and negative $C_{11}$, i.e., $v_l = \sqrt{C_{11}/r_{11}}$ is imaginary. That is to say, the third negative band (T) can only hold the transverse wave propagation. We also notice that the performance of the effective bulk modulus $K$ is different from those of $C_{11}$ and $C_{22}$, which implies the strong anisotropy of the optimized EMM. We finally find that all the effective moduli in Fig. 5(c) achieve the extrema at two frequencies of 33.95 kHz and 43.87 kHz. That is, all associated applied displacement fields will stimulate the resonances of the optimized EMM. This declares that the optimized EMM has very complex effective motions at the high frequencies, and all types of vibrations are coupled with each other in the modes at the lower edges of the sixth and eighth bands. We stress that the range of double-negative $\rho_{11}$ and $C_{11}$ is in good agreement with the negative band (L) no matter how complicated the EMM is. This clarifies the validity of our optimization formulation and results. Unlike the four-phase EMM reported by Lai et al. (2011), three negative bands (L, qT and T) in Fig. 5(a) have the broadband negative index. Similar to their reported band characteristics (Lai et al., 2011), the band L can only support the longitudinal wave; and the bands qT and T allow the transverse wave. However, the corresponding physical mechanisms shown in the next section are quietly different from their findings (Lai et al., 2011).

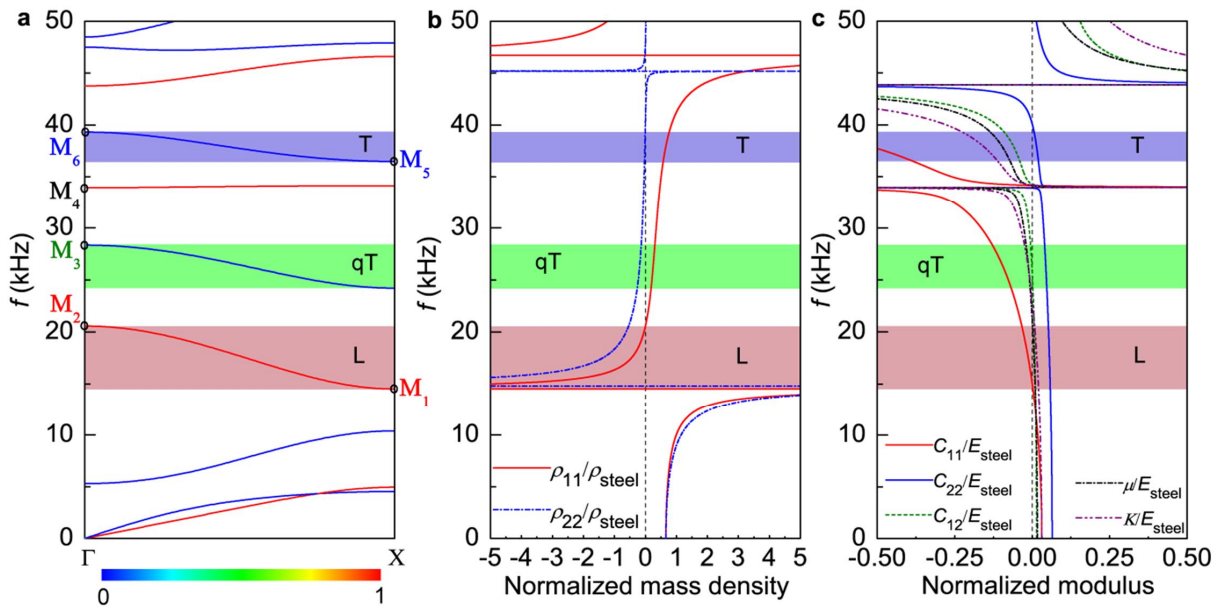

**Fig. 5.** Dispersion curves and effective material parameters for the EMM in Fig. 4(a). **(a)** The band structure along ΓX direction for the in-plane waves (the normalized quantity $q_x$ associated with the displacement $u_x$ component is displayed by the corresponding color); **(b)** effective mass density along $x$ (red solid line) and $y$ (blue dash line) directions; and **(c)** effective elastic modulus ($C_{11}$: red hollow circle; $C_{22}$: bule hollow upper triangle; $C_{12}$: green hollow square; $\mu$: black hollow rotational square and $K$: purple hollow lower triangle). The frequency ranges of three negative bands in (a) supporting the longitudinal (L), quasi-transverse (qT) and transverse (T) waves are marked by the pink, green and blue bars, respectively.

To further confirm the unusual wave properties, we investigate the transmission of waves along ΓX direction through a finite sample based on the optimized EMM in Fig. 4(a), see Fig. 6. We consider the input longitudinal (Fig. 6(a)) and transverse (Fig. 6(b)) displacement excitations and collect the output responses. The periodic boundary conditions are added on the upper and lower edges. And the solids on the left of the input line and on the right of the output line are the infinite elements (CINPE4 in ABAQUS) which act as the absorbing boundaries. It is clearly seen that, under the longitudinal input, large transmissions for the longitudinal wave are obtained in the first negative band (L). In contrast, under the transverse input, large transmissions for the transverse wave are obtained in the second (qT) and third (T) negative bands. Due to the same incident transverse wave, it is not possible to distinguish the quasi-transverse and transverse waves from the transmission calculations. In view of the large difference in the order of the magnitude, the transmitted waves can be regarded as the pure waves. Obviously, these results are exactly agreement with those predicted by the dispersion curve and effective medium theory.

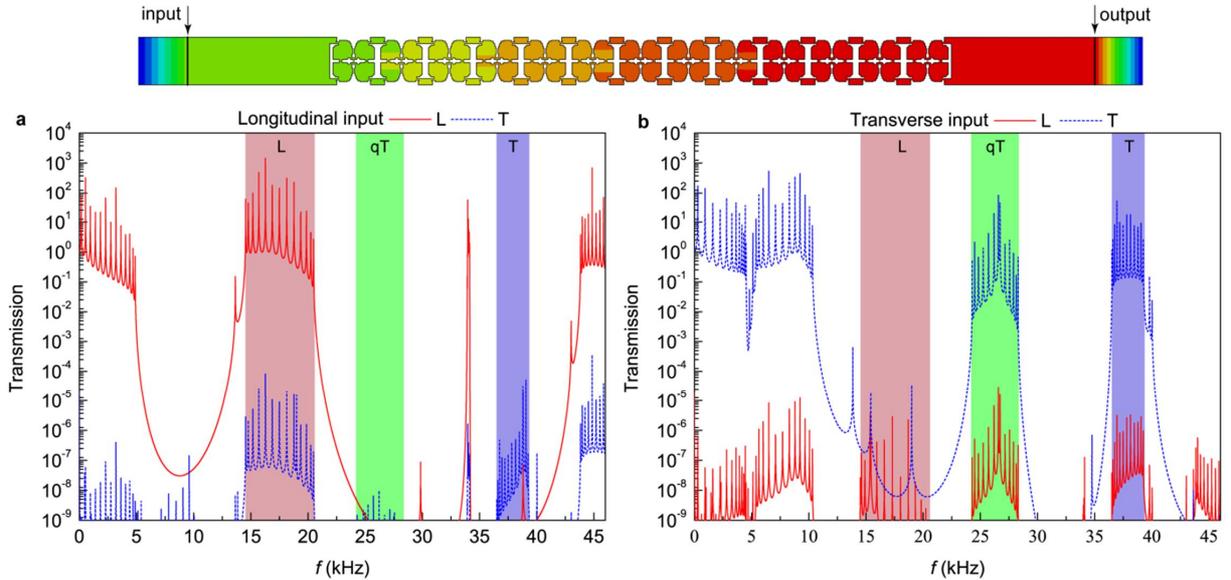

**Fig. 6.** Transmission through a finite sample based on the EMM in Fig. 4(a). **(a)** Transmission along ΓX direction for the longitudinal input excitations and **(b)** the transverse input excitations. The obtained longitudinal wave transmission and transverse wave transmission are marked by the red solid and blue dashed lines, respectively. The finite sample for transmission computation is presented in the top. With the incident wave added on the input, the response at the output is calculated. The solids on the left of input line and on the right of output line are the infinite elements (CINPE4 in ABAQUS) which act as the absorbing boundaries.

To understand how the incident wave propagates through the EMM in an arbitrary direction, we plot in Figs. 7(a) and 7(b) the eigenfrequency curves (EFCs) of material in the frequency ranges of 14-21 kHz and 34.5-39.5 kHz in the first Brillouin zone. It is noted that, both the first and third negative bands in Fig. 5(a) are quite anisotropic. The anisotropy becomes weak as the frequency increases with the Bloch wave vectors decreasing. The variation of the EFCs in Fig. 7(a) shows that, with frequency increasing, the negative group velocities for the longitudinal wave occur along ΓX direction. Likewise, the EFCs in Fig. 7(b) show the existence of negative group velocities for the transverse wave in ΓX direction. Therefore, negative refraction of both longitudinal and transverse waves is expected and will be demonstrated and discussed in Sec. 4.4.1.

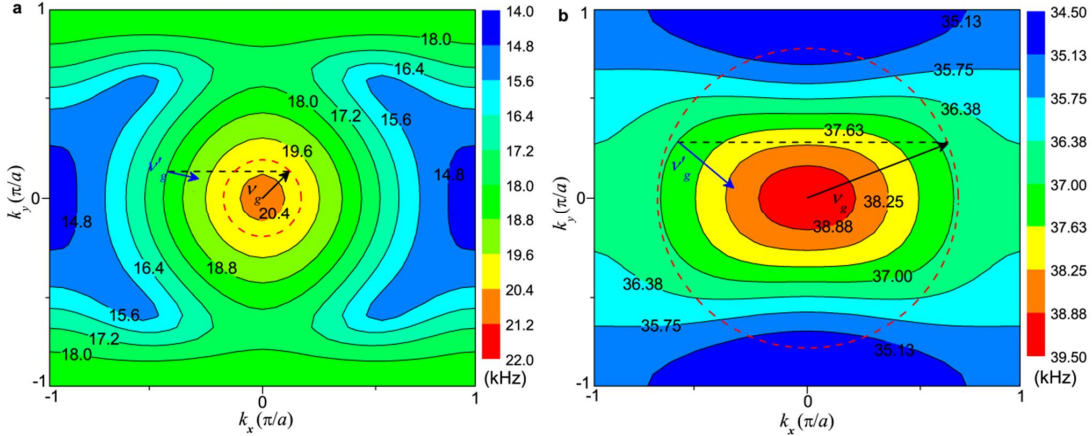

**Fig. 7. EFCs for the longitudinal and transverse wave modes. (a)** EFCs of the forth band in Fig. 5(a); **(b)** EFCs of the seventh band in Fig. 5(a).

To explore the effect of the wavelength scale, we also perform optimization under two upper operating frequencies $f_{max}$: 9.75 and 39.04 kHz, see Fig. 8. Their associated wavelengths are 20 and 5 times larger than the lattice constant $a$, respectively. The corresponding double-negative ranges are 7.44-10.02 kHz (Fig. 8(a)) and 32.8-46.13 kHz (Fig. 8(c)), respectively, which are highly consistent with the negative band ranges. The optimized structure of Fig. 4(a) with the double-negative range 0f 15.09-20.57 kHz for $f_{max}$=19.5 kHz is also presented for comparison (Fig. 8(b)). As for the optimized geometry, three EMMs show the similar topological features, i.e., four big central lumps are connected with the other four corner blocks. It is noted that, with the increase of the target frequency range (larger $f_{max}$), the corner blocks become bigger and bigger. Specially, we find that their relative bandwidths are 0.295, 0.307 and 0.338, respectively. This means that, with the same constraints, our proposed optimization method can find out the optimized structures with similar resonances whose relative bandwidths are nearly the same. On the other hand, it is easy to design the expected double-negative bandwidth by only changing the upper operating frequency $f_{max}$. This is likely to be very useful in designing the negative-index metamaterials.

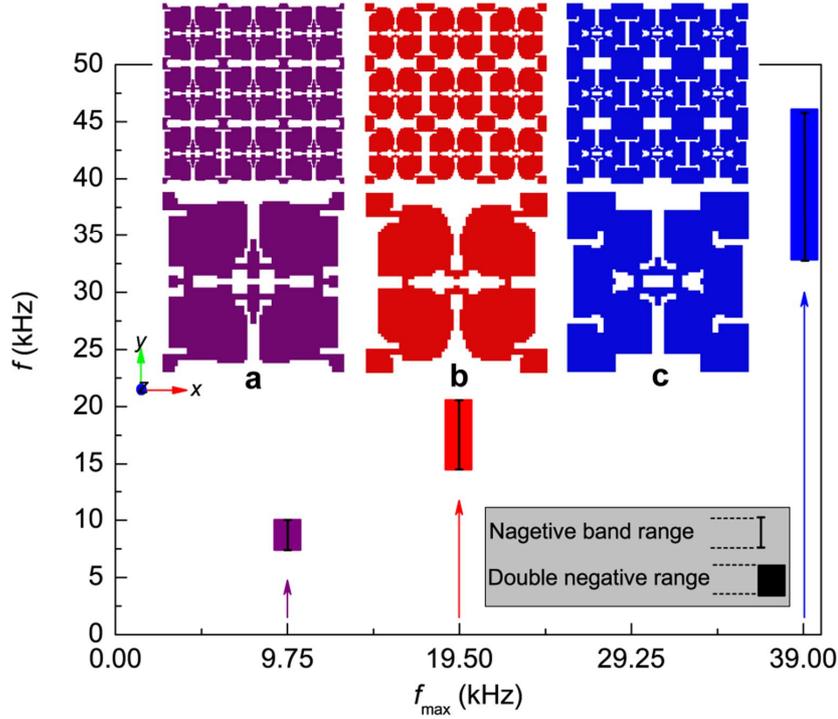

Fig. 8. Optimized EMMs for the different target frequencies ($f_{min}$=0.5 Hz, $\alpha$=0.5, $\beta$=0.5 and $\delta_C$=0.3). (a) $f_{max}$=9.75 kHz; (b) $f_{max}$=19.5 kHz; and (c) $f_{max}$=39.04 kHz.

### 4.2. Mechanism of double-negativity

To understand the mechanism of the aforementioned features, the eigenstates of the lower and upper edges (M1-M6 in Fig. 5(a)) are illustrated in Fig. 9. Overall the low symmetry of the unitcell results in the strong anisotropy for the dynamic responses. For eigenstate $M_1$, the energy mostly concentrates in the four central lumps; the four corner blocks can hardly vibrate. Intuitively, this quadrupolar rotational resonance can produce the negative $\rho_{11}$. With a smaller wave vector, the eigenstate $M_2$ has translation in four small corner blocks and rotation in four big central lumps. Both eigenstates $M_1$ and $M_2$ have the rotation in the upper two lumps and the opposite rotation in the lower two. We believe that it is the change of the rotating center that causes the different negativity extent of $\rho_{11}$ at different frequencies. Obviously, the boundary response of mode $M_2$ is a typical longitudinal wave motion. This, in turn, explains why $\rho_{11}$ in Fig. 5(b) become negative within the band L. So, over the whole band region, the boundary vibration along *x*-direction proves that the first negative band (L) indeed supports the longitudinal wave propagation. Interestingly, for the eigenstate $M_3$, the rotation in four big central lumps induces apparent opposite rotation in four small corner blocks, indicating the transverse wave motion. Nevertheless, the distribution of the displacement component $u_y$ in the boundary nodes attests that the whole effective motion of $M_3$ is not a purely transverse wave, but a quasi-transverse wave motion. As shown in Fig. 5(c), $\mu$ and $C_{12}$ have the same tendency as the frequency arises. In other words, the negative $C_{12}$ is often along with the occurrence of negative $\mu$. In fact, for the orthogonal-symmetric structures in this paper, $C_{12}$ is easy to become negative when $C_{11}$ is negative. The property of possessing the simultaneous negative $C_{12}$ and $\mu$ usually produces the quasi-transverse wave band. Aiming at the pure longitudinal wave motion of the optimized EMM within the target frequency range, we have to separate the longitudinal and transverse wave by using the constraint in Eq. (16). Otherwise, the first and second negative bands will be coupled together, making the eigenstates very complex, and thus no pure wave mode can be obtained. In particular, the eigenstate $M_4$ shows the negative longitudinal modulus: the small corner blocks vibrate slightly; the big central lumps show obvious opposite rotations. It is not difficult to image the simultaneous negative $C_{11}$ and $K$ from the mode $M_4$. We also view that the boundary responses in the modes $M_5$ and $M_6$ respectively show up the shear deformation along *x*-direction and translation motion along *y*-direction under the different rotation and vibration combination. Unlike the modes $M_1$-$M_4$, the eigenstate $M_5$ concentrates the most energy in the four corner blocks. Their vibrations force the big central lumps to translate along different directions. The whole boundary effective motion is completely the shear deformation. As for the eigenstate $M_6$, the translation motion of four solid corner blocks makes the central lumps rotate. The corresponding boundary displacement field implies the pure transverse wave motion along *x*-direction. Without any rotation in the corner solid blocks, the whole effective vibrations of $M_5$ and $M_6$ can be regarded as purely transverse wave motions. This means that the third negative band (T) only holds the transverse wave along ΓX direction. Therefore, the optimized orthogonal EMMs in this paper are very easy to form the purely transverse wave band with the negative index at high frequencies, even though we just focus on the optimization about the simultaneous negative $\rho_{11}$ and $C_{11}$ at the deep-subwavelength scale.

Apparently, the similar rotations of quadrupolar resonances associated with the negative mass density and longitudinal wave modulus is the key factor to help the optimized EMM to generate the broadband longitudinal wave band with the negative index. The following eigenstate analysis of the optimized low frequency EMM (Fig. 8(a)) and high-frequency one (Fig. 8(c)) will further show that the multipolar resonance is a key factor to yield double-negativity. Figures 10(a) and 10(b) display the eigenstates for negative $C_{11}$ and negative $r_{11}$ of the optimized low-frequency EMM, respectively; Figures 10(c) and 10(d) display the eigenstates for negative $C_{11}$ and negative $r_{11}$ of the optimized high-frequency EMM, respectively. Unlike the quadrupolar resonances for negative $C_{11}$ and $\rho_{11}$ in Fig. 9, the multipolar resonances are demonstrated. The high-frequency vibrations (Figs. 10(a) and 10(b)) are generated by the rotation of four lumps and translation of two blocks. So the thicker connections will result in the higher frequency. Compared with the eigenstates $M_4$ and $M_1$ in Fig. 9, the low-frequency vibration adds four resonators which make the operating frequency very low. In fact, the four new small solid blocks can help the four big lumps to induce rotations more easily, leading to the generation of the low-frequency negative longitudinal band. Therefore, differing from the reported four-phase metamaterials (Lai et al., 2011), the multipolar resonances in Fig. 10 provide more possibilities to lower or raise the frequency.

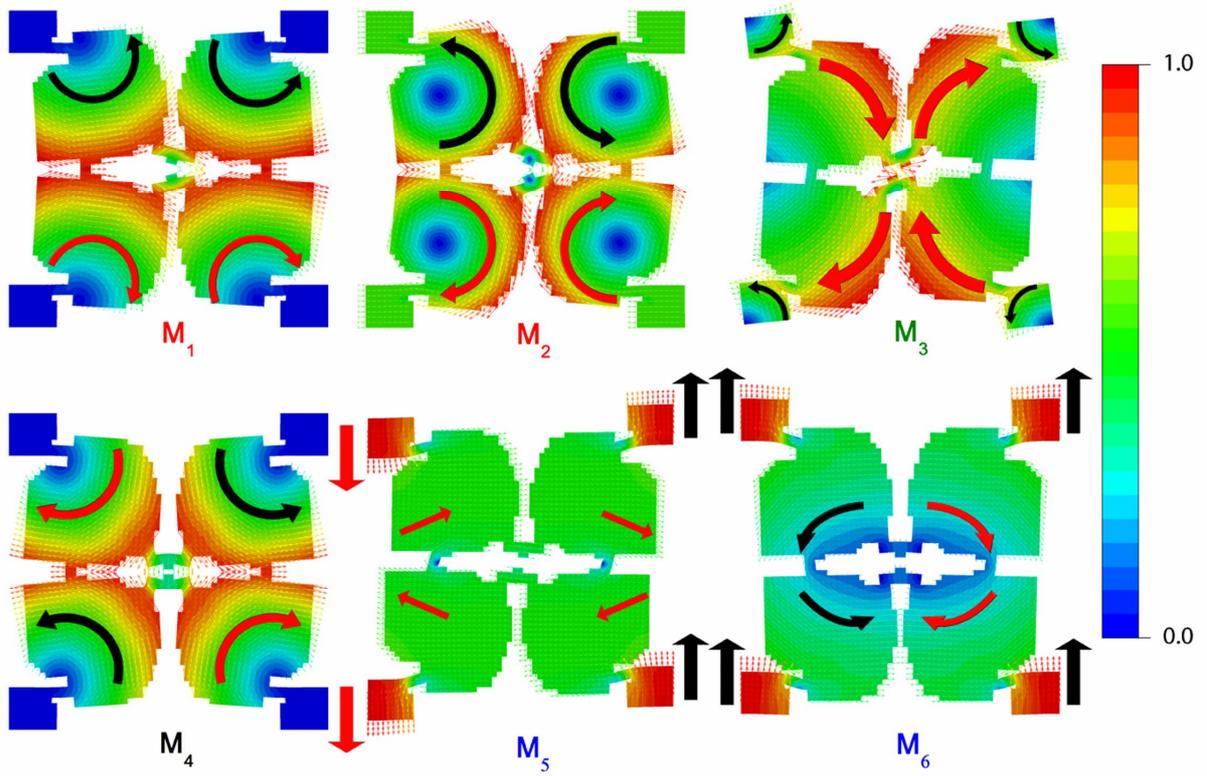

**Fig. 9.** Field distributions of eigenstates marked in Fig. 5(a).

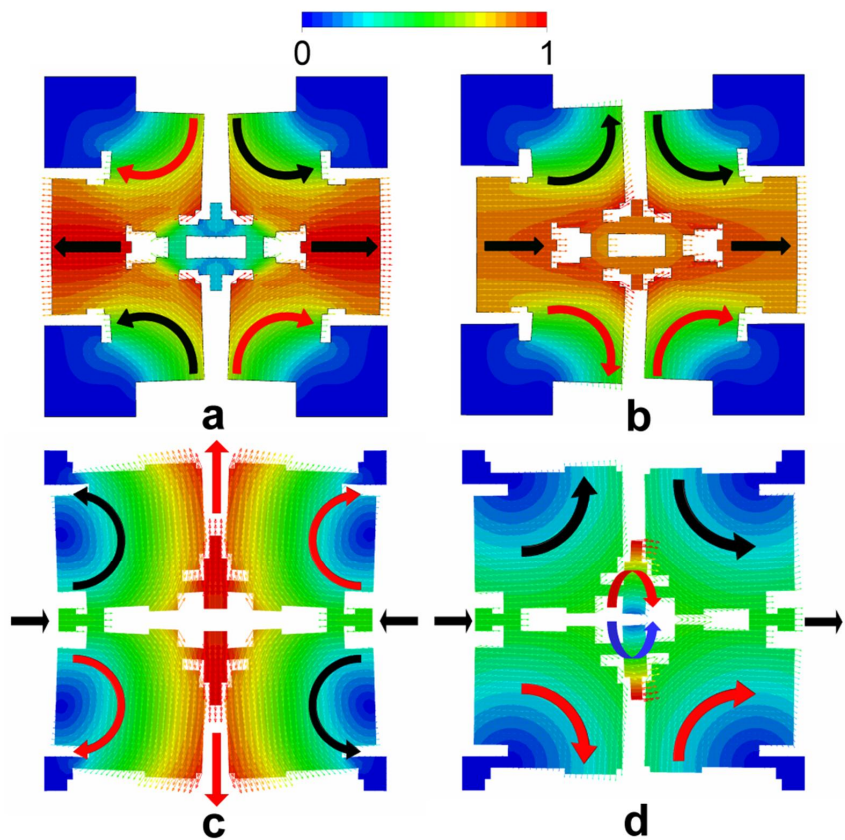

**Fig. 10.** Mechanisms for optimized EMMs in Figs. 8(c) and 8(a). **(a)** Eigenstate with frequency of 51599 Hz and $k_x$=0 for negative $C_{11}$ of the EMM in Fig. 8(c); **(b)** eigenstate with frequency of 32743 Hz and $k_x$=π/$a$ for negative $\rho_{11}$ of the EMM

in Fig. 8(c); **(c)** eigenstate with frequency of 30937 Hz and $k_x=0$ for negative $C_{11}$ of the EMM in Fig. 8(a); and **(d)** eigenstate with frequency of 7439.7 Hz and $k_x=\pi/a$ for negative $\rho_{11}$ of the EMM in Fig. 8(a);

### 4.3 Influences of the optimization parameters

It is noticed that several parameters are involved in the optimization formulation (13)-(17). Selection of the values of these optimization parameters, especially $d_C$, $\alpha$ and $\beta$, will influence the optimized structures. In this section we will discuss the effects of $d_C$, $\alpha$ and $\beta$ to give an insight into double-negative index optimization.

### 4.3.1 Influence of $d_C$

The value of $\delta_C$ in **Eq. (16)** reflects the influence of $C_{12}$ on the optimized results. To demonstrate the effect of $d_C$ or $C_{12}$ on optimization, we design two EMMs under the constraints of $\delta_C=0.2$ and $\delta_C=0.42$ with the target frequency $f_{max}=19.5$ kHz, as shown in **Figs. 11(a)** and **11(b)**, respectively. Clearly, the biggest difference between these two EMMs is that the stronger connections appear between the upper and lower central solid lumps in **Fig. 11(b)**. This restrictive feature makes the EMM more difficult to generate coupling along *y*-direction. Then, the energy mainly propagates along *x*-direction. Besides, the four corner solid blocks in **Fig. 11(b)** are smaller than those in **Figs. 4(a)** and **11(a)**. As a result, a wider double-negative index range and larger refraction angle are achieved; see comparison of the corresponding EFCs illustrated in **Figs. 11(c)** and **11(d)**. It is seen that the frequency increases as the wave vector gets smaller. So the curve shapes in **Figs. 11(c)** and **11(d)** mean that the negative group velocity can be generated along ΓX direction. For these two optimized EMMs, the detailed ranges of the negative group velocity along ΓX direction and the ranges of the double negative index are presented in **Table 1** (see the first three rows). Clearly, two ranges coincide with each other. The largest amplitude of the negative group velocity occurs near the mid-frequency of the range. As $\delta_C$ increases, the EMM has a wider negative band range and a larger amplitude of the negative group velocity. A large amplitude of the group velocity generally means the high energy transmission, and results in a relatively high imaging resolution. Therefore, changing the coupling stiffness $C_{12}$ (or $\delta_C$) is very useful to design the EMM having a broad negative band and a large refraction angle.

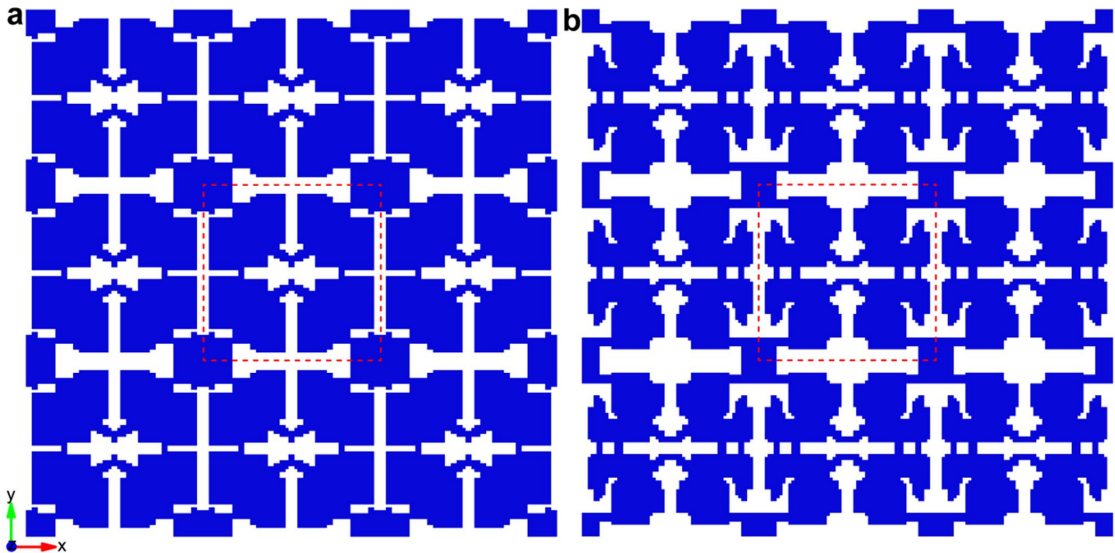

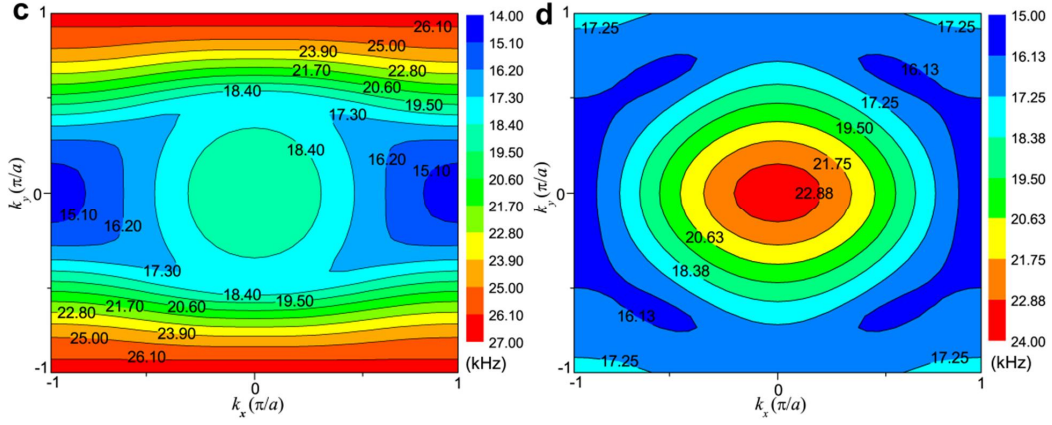

Fig. 11. Optimized EMMs with the different $C_{12}$ constraints ($f_{min}$=0.5 Hz, $f_{max}$=19.5 kHz, $\alpha$=0.5 and $\beta$=0.5). (a) The EMM with $\delta_C$=0.2; (b) EMM with $\delta_C$=0.42; (c) EFCs of the forth band with $\delta_C$=0.2; and (d) EFCs of the forth band with $\delta_C$=0.42.

### 4.3.2 Influences of $\alpha$ and $\beta$

Given a large search space of optimization involving two objectives, we next explore the role of the weight coefficients, $\alpha$ and $\beta$, in optimization by considering the three different combinations (see Figs. 12(a)-12(c)) and comparing the results with the EMM in Fig. 4(a) ($\alpha$=0.5, $\beta$=0.5). The structures in Figs. 12(a) ($\alpha$=0.4, $\beta$=0.6) and 12(b) ($\alpha$=0.6, $\beta$=0.4) show the topological features similar to Fig. 4(a). The main difference is the geometries of the solid lumps and locations of the connections. However, the EMM in Fig. 12(c) ($\alpha$=0.7, $\beta$=0.3) has more complex geometry, i.e., more solid blocks and completely different connections. The corresponding EFCs of these three combinations are given in Figs. 12(d), 12(e) and 12(f), respectively. It is noted that, with the increase of $\alpha$, the curves in ΓY direction get bigger changes. Roughly speaking, the curvatures of EFCs in Fig. 12(e) are larger than those in Fig. 12(d). Even the curvature in the opposite direction is obtained (Fig. 12(f)). A large $\alpha$ usually induces the EFCs to generate negative group velocities in large $k_y$ ranges. In fact, the EMM with highly symmetrical solid blocks combined with narrow connections is conducive to the rotation like eigenstate $M_4$ (see Fig. 9), which can induce the negative $C_{11}$. At the same time, the rotation like eigenstates $M_1$ and $M_2$ (see Fig. 9), and thus the negative $\rho_{11}$ is relatively difficult to be generated. The widths of the negative group velocity ranges for four combinations are 5.19 ($\alpha$=0.4, $\beta$=0.6), 6.05 ($\alpha$=0.5, $\beta$=0.5), 4.25 ($\alpha$=0.6, $\beta$=0.4), and 5.67 ($\alpha$=0.7, $\beta$=0.3) kHz, see Table 1 (see the second and last three rows). The double-negative index ranges match the negative group velocity ranges well. For the frequencies below $f_{max}$ (19.5 kHz), the cases with the combinations of $\alpha$=0.4, $\beta$=0.6 and $\alpha$=0.7, $\beta$=0.3 have the widths of 3.73 and 3.75 kHz for the negative group velocity ranges, respectively. But the EMM with $\alpha$=0.6, $\beta$=0.4 produces a larger width of 4.26 kHz. Since the combinations of $\alpha$=0.4, $\beta$=0.6 and $\alpha$=0.6, $\beta$=0.4 have the same bias from $\alpha$=0.5, $\beta$=0.5, their difference in the width below $f_{max}$ shows that the preference to $\rho_{11}$ is beneficial to widen a negative band. In general, we can easily control the proportion of two resonant modes of the optimized EMM for the negative $\rho_{11}$ and $C_{11}$ by setting different weight coefficients of the objective functions. Anyway, for the double-negative index optimization, we suggest that it is better to treat the mass density and modulus equally.

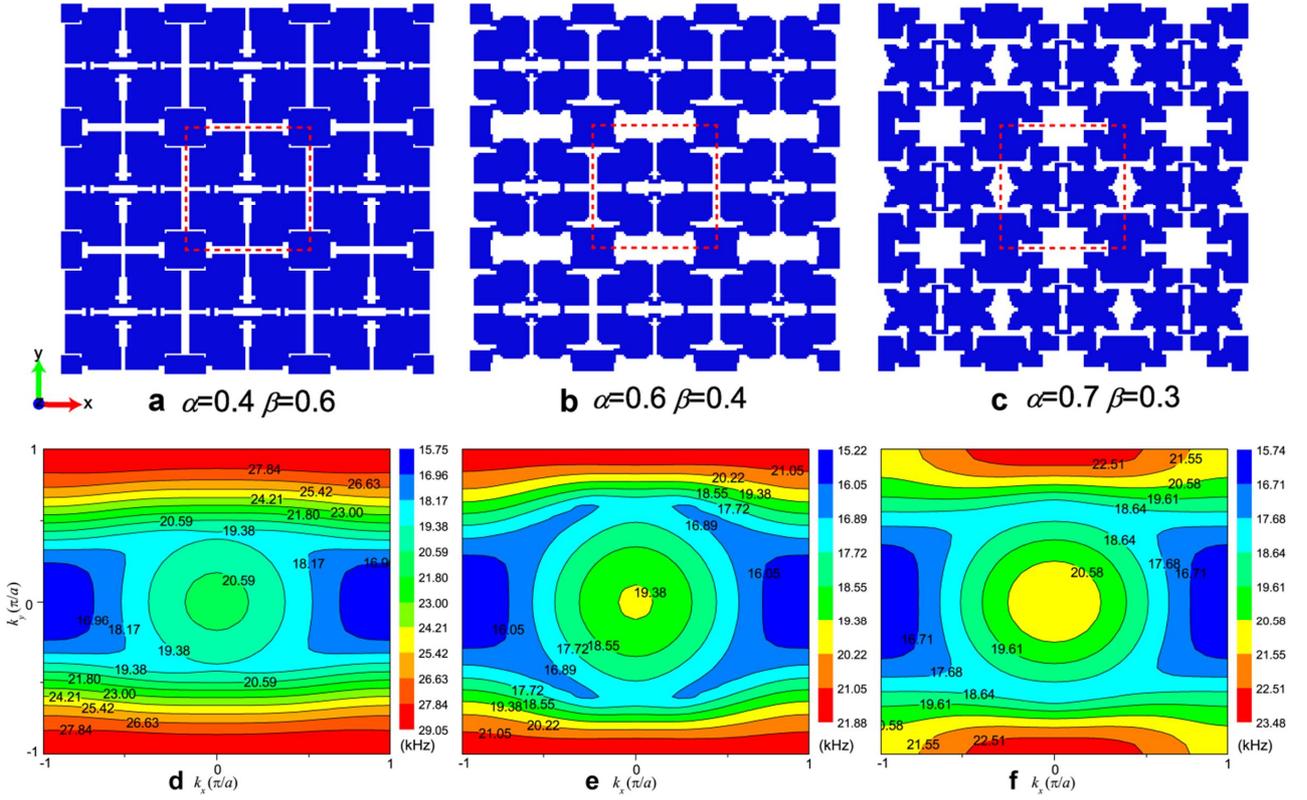

**Fig. 12.** Optimized EMMs with the different coefficients of objectives ($f_{min}$=0.5 Hz, $f_{max}$=19.5 kHz and $\delta_C$=0.3). **(a)** The EMM with $\alpha$=0.4, $\beta$=0.6; **(b)** the EMM with $\alpha$=0.6, $\beta$=0.4; **(c)** the EMM with $\alpha$=0.7, $\beta$=0.3; **(d)** EFCs of the forth band for EMM in (a); **(e)** EFCs of the forth band for EMM in (b); and **(f)** EFCs of the forth band for EMM in (c).

**Table 1**

Comparison of ranges for the negative group velocity along ΓX direction and double-negative parameters ($\rho_{11}$ and $C_{11}$).

| Parameter combination | Double-negative range (kHz) | Negative group velocity range (kHz) | Minimal group velocity (m/s) |
|---|---|---|---|
| $\alpha$=0.50 $\beta$=0.50 $\delta_C$=0.20 | 15.55-19.50 | 14.69-19.49 | -439.2 @ 16.94 kHz |
| $\alpha$=0.50 $\beta$=0.50 $\delta_C$=0.30 | 15.09-20.57 | 14.52-20.57 | -556.2 @ 17.36 kHz |
| $\alpha$=0.50 $\beta$=0.50 $\delta_C$=0.42 | 15.66-23.57 | 15.60-23.55 | -702.0 @ 18.41 kHz |
| $\alpha$=0.40 $\beta$=0.60 $\delta_C$=0.30 | 16.07-20.97 | 15.77-20.96 | -480.0 @ 18.14 kHz |
| $\alpha$=0.60 $\beta$=0.40 $\delta_C$=0.30 | 15.60-19.50 | 15.23-19.48 | -393.6 @ 17.21 kHz |
| $\alpha$=0.70 $\beta$=0.30 $\delta_C$=0.30 | 16.42-21.45 | 15.75-21.42 | -493.2 @ 17.97 kHz |

### 4.4. Anomalous elastic wave properties and emerging applications

The above optimized EMMs exhibit newfangled physical properties, such as anisotropic mass density, negative mass density, negative elastic moduli, negative group velocity and their various positive-negative combinations. These anomalous features imply prospective applications of the EMM. in manipulating elastic waves in solids including the negative refraction, imaging, and cloaking effect, etc. This section shows some emerging applications of the optimized EMMs.

#### 4.4.1. Negative refraction and imaging

All physical quantities, including the dispersion relation, normalized quantity $q_x$, effective material parameters and EFCs, can predict that the optimized structure in Fig. 4(a) can lead to the negative refraction of the longitudinal and transverse waves within certain frequency ranges along ΓX direction. According to Fig. 7(a), the refracted waves will propagate along the negative direction when a longitudinal wave is incident to the interface between a homogeneous solid and the EMM. Similarly, from Fig. 7(b), the negative refraction phenomenon also can be observed if a transverse wave is incident. Their negative properties can be easily understood by comparing the directions of group velocities of the incident waves ($v_g$) and refracted waves ($v'_g$), see Fig. 7 where the red dashed circles are the EFCs of the longitudinal wave in stainless steel at 18 kHz (Fig. 7(a)) and the transverse wave at 37 kHz (Fig. 7(b)), respectively.

For validating the negative refraction, the displacement field patterns of the longitudinal wave component under the incident longitudinal wave at 18 kHz are shown in Fig. 13(a). Clearly, the refracted waves in the EMM propagate along the negative direction, and the distinct negative refraction appears in the right part of the model. It is well known that the negative refraction property is helpful in design of a flat lens. Figure 13(b) presents the imaging field pattern of the longitudinal wave component. We can notice the apparent subwavelength imaging with the full width at the half maximum (FWHM) of image being 0.465$\lambda$ which breaks the diffraction limit of 0.5$\lambda$. This can be understood by considering the reconstruction of the details of the object through the amplification of the evanescent waves (Christensen and García de Abajo, 2012). Furthermore, the displacement field patterns of the transverse wave component under the incident transverse wave at 37 kHz are displayed in Fig. 13(c) which shows obvious negative refraction. Even more momentously, negative refraction of a transverse wave in the single-phase EMM has never been realized before. The negative fraction and imaging phenomena demonstrate well the correctness of our optimization results.

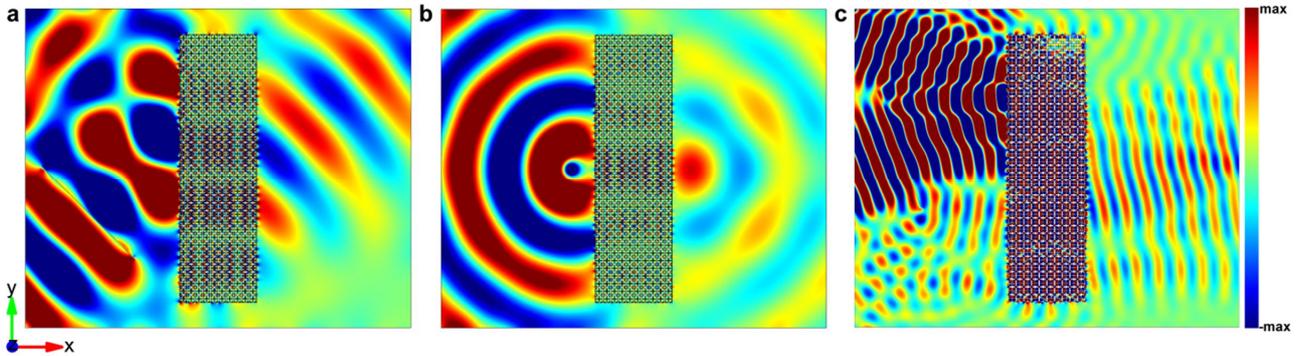

Fig. 13. Negative refraction and imaging of the EMM in Fig. 4(a). (a) The displacement field patterns of longitudinal wave component generated by an incident Gaussian beam (45°) of longitudinal wave at 18 kHz; (b) the imaging field patterns (the source is located in the position 2$a$ away from the left side of slab) of longitudinal wave component at 18 kHz (FWHM=0.465$\lambda$); and (c) the displacement field patterns of transverse wave component generated by an incident Gaussian beam (20°) of transverse wave at 37 kHz. Note that, for all wave simulations in this paper, the EMM slabs are surrounded by stainless steel.

Because of the relative small target wavelength ($\lambda=5a$), the optimized EMM in Fig. 8(c) is expected to make the large attenuation of the evanescent waves for imaging. We present the negative refraction and imaging magnitude fields with FWHM=0.696$\lambda$ at 39 kHz in Figs. 14(a) and **14(b)**, respectively. Compared with the negative refraction in Fig. 13(a), the phenomenon in Figs. 14(a) shows the larger angle of refraction and higher transmission. In fact, the relative effective index of refraction of the EMM at a high frequency is bigger, resulting in the larger angle of refraction. In the present case, we argue that relative high transmission is due to the better matching impedance and relatively low resonance coupled with the traveling modes of the EMM, implying more energy through the metamaterials. Consequently, the image has the large magnitude of the longitudinal wave component, see Fig. 14(b). In Fig. 14(c), we show the focal length from the far side of the EMM slab to the image and FWHM of the image spot as the functions of the frequency (37-39.5 kHz). Obviously, a larger operating frequency results in the farther image. Nevertheless, we note that FWHM is not monotonically increasing with the frequency. A dip occurs near 38.75 kHz. After this dip, the FWHM has a rapid increase as the frequency increases. Generally, the evanescent waves decay sharply when they are far away from the interface. The usual FWHM change is caused by the irregular transmittance of the evanescent waves. Obviously, unlike the imaging in Fig. 12(b), the EMM with a relative small wavelength does not yield a diffraction-limit resolution. This emphasizes the key role of the wavelength scale in EMM imaging.

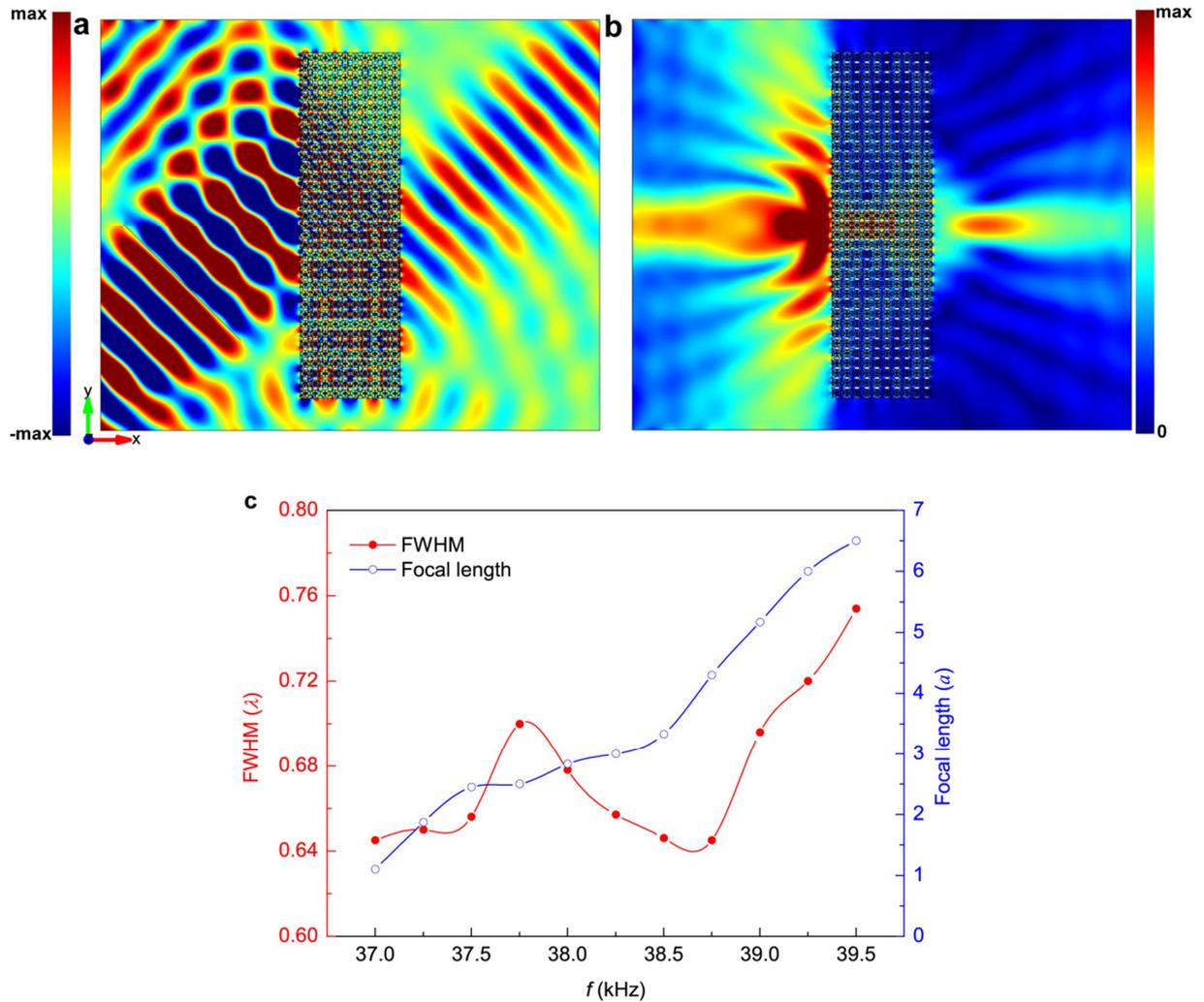

**Fig. 14.** Negative refraction and imaging of the EMM in Fig. 8(c). **(a)** The displacement field patterns of longitudinal wave component generated by an incident Gaussian beam (45°) of longitudinal wave at 39 kHz; **(b)** the imaging magnitude field patterns (the source is located in the position 2$a$ away from the left side of slab) of longitudinal wave

component at 39 kHz (FWHM=0.696$\lambda$); and **(c)** focal length from the far side of the EMM slab to the image (line with hollow circle) and FWHM of the image spot (line with solid circle) as a function of frequency.

### 4.4.2 Near-zero mass density and cloaking effect

As shown in **Figs. 5(b)** and **5(c)**, remarkably, the optimized EMM in **Fig. 4(a)** can be regarded as an anisotropic zero-index metamaterial (ZIM) along *x*-direction at the operating frequency where the near-zero $\rho_{11}$ is induced. During the last decade, ZIMs, which have simultaneously or individually near-zero effective material parameters, have aroused great interest among the researchers because they show many intriguing phenomena and applications for electromagnetic waves such as electromagnetic and acoustic cloaking (**Hao et al., 2010**; **Zhang and Wu 2015**), tunneling of energy through bent channels (**Silveirinha and Engheta 2006**), directional radiation patterns (**Alù et al., 2007**) and zero phase accumulation (**Mahmoud and Engheta 2014**). On account of the complexity of elastic wave scattering, relatively few works were reported about ZIMs for elastic waves (**Liu et al., 2011**; **Antonakakis et al., 2014**; **Liu and Liu 2015**).

Here, we focus the frequency of 20.5184 kHz, where $\rho_{11}/\rho_{steel}$=-0.004→0⁻, $\rho_{22}/\rho_{steel}$=-0.55, $C_{11}/E_{steel}$=-0.033, $C_{22}/E_{steel}$=0.052 and $\mu/E_{steel}$>0. These effective parameters mean that only the longitudinal wave ($\sqrt{C_{11}/\rho_{11}} > 0$) along *x*-direction can propagate in the EMM while the transverse wave ($\sqrt{\mu/\rho_{22}} < 0$) along *x*-direction are prohibited. At the same time, both the longitudinal ($\sqrt{C_{22}/\rho_{22}} < 0$) and transverse ($\sqrt{\mu/\rho_{11}} < 0$) waves along *y*-direction cannot pass through the EMM. Unlike the reported mechanisms (simultaneous near-zero $\rho$ and $1/\mu$) by **Liu and Liu** (**2015**), our optimized EMM shows high anisotropy. However, we find some common features: near-zero mass density and the ability of no shear deformation during wave propagation. Accordingly, we believe that our optimized EMM is indeed a ZIM along *x*-direction. **Figure 15(a)** illustrates the wave transmission through a solid (stainless steel) strip filled with the optimized EMM shown in **Fig. 4(a)**. A rectangular hole with the size of 5*a*×3*a* is introduced in the middle of the structure. A plane longitudinal wave at 20.5184 kHz is incident from the left. Because of the nearly infinite phase velocity in the EMM along *x*-direction, the phase of the wave throughout the whole EMM is nearly constant. In addition, the longitudinal polarization along *x*-direction completely dominates the displacement field. As a result, the so-called cloaking effect can be observed from **Fig. 15(a)**. For comparison, we plot in **Fig. 15(b)** the displacement pattern for the sample without the optimized EMM inside. Unsurprisingly, the strong scattered waves are excited by the rectangular hole, resulting in obvious distortion of the wave fronts.

Interestingly, we find the similar effective material features for the transverse wave at 39.2262 kHz with $\rho_{22}/\rho_{steel}$=-0.008→0⁻, $\rho_{11}/\rho_{steel}$=0.758, $C_{22}/E_{steel}$=0.0069, $C_{11}/E_{steel}$=-0.7121 and $\mu/E_{steel}$=-0.1321. Obviously, only the transverse wave ($\sqrt{\mu/\rho_{22}} > 0$) along *x*-direction can propagate through the optimized EMM. Therefore, we believe that the EMM is also a ZIM for the transverse wave along *x*-direction. As expected, **Fig. 15(c)** shows the exciting cloaking effect for the transverse wave as well, totally overcoming the scattering of waves as shown in **Fig. 15(d)**.

It is necessary to note that the cloaking effect is very sensitive to the operating frequency (**Liu and Liu 2015**; **Zhang and Wu 2015**). In other words, a small change of the frequency will cause the reduction of the transmission efficiency. We stress that most optimized double-negative EMMs in this paper can serve as anisotropic ZIMs along *x*-direction at a certain frequency, where only the longitudinal wave along *x*-direction can be supported when negative near-zero $\rho_{11}$ is obtained. In addition, these optimized EMMs can also be treated as the ZIMs for the transverse wave at a high frequency where the negative near-zero $\rho_{22}$ is achieved.

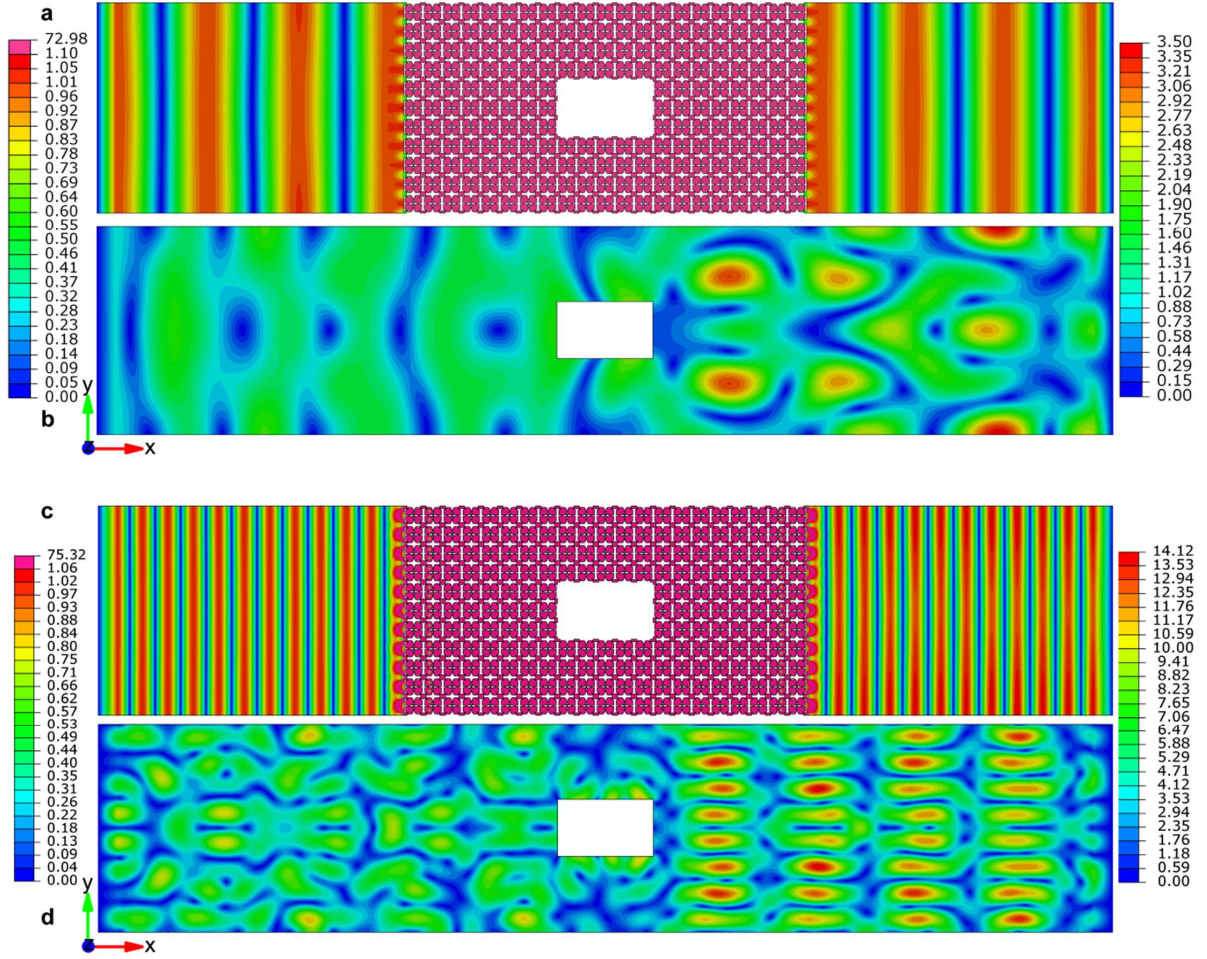

**Fig. 15.** Wave transmission through a stainless steel waveguide filled with/without the optimized EMM in Fig. 4(a). **(a)** Displacement field pattern ($\sqrt{u_x^2+u_y^2}$) for a longitudinal plane wave at 20.5184 kHz incident from the left side of the waveguide with the EMM; **(b)** result without the EMM under the same excitations in (a); **(c)** Displacement field pattern ($\sqrt{u_x^2+u_y^2}$) for a transverse plane wave at 39.2262 kHz incident from the left side of the waveguide with the EMM; and **(d)** result without the EMM under the same excitations in (c). The same rectangle defect with the size of 5*a*×3*a* is introduced in the center of the waveguide. The upper and lower left scale bars show the amplitudes of the displacement field in (a) and (c), respectively. The upper and lower right scale bars show the amplitudes of the displacement field in (b) and (d), respectively. For all simulations, the periodic boundary conditions are assumed on the upper and lower boundaries, and the infinite elements (CINPE4 in ABAQUS) are added on the left and right boundaries to avoid wave reflection.

## 5. Optimized super-anisotropic EMM with hyperbolic dispersion

From all optimized EMMs in Sec. 4, we find a very interesting phenomenon, i.e., the range of negative $C_{11}$ is much larger than that of negative $\rho_{11}$. Generally, we find that if the effective moduli are negative in a frequency range ($f_1$, $f_3$), then the mass densities are negative in a range of ($f_1$, $f_2$) with $f_2<f_3$. That is to say, ($f_1$, $f_2$) is a double negative range, while ($f_2$, $f_3$) is a single negative range with positive mass densities and negative moduli. This implies that the elastic waves show the evanescent property and cannot propagate in the EMM within the range ($f_2$, $f_3$) along one principle direction. In this case, if waves can propagate along other directions, a typical hyperbolic dispersion is obtained (Christensen and García de Abajo, 2012; Lee et al., 2016). Now that the EMMs presented in this paper are very easy to produce a simultaneous negative $\rho_{11}$ and a negative $C_{11}$ below a certain frequency, can we design positive $\rho_{11}$ and negative $C_{11}$ in a range above this frequency with both $\rho_{22}$ and $C_{22}$ being positive? To achieve this goal, we revise the objective function in Eq. (14) and the constraints as:

$$Maximize: D' = ND' + \min\left\{-0.5 \times \frac{\min_{\forall m'(m' \in (1,2,\ldots M'))}\left[\rho_{11}^+(m')\right]}{\max_{\forall m'(m' \in (1,2,\ldots M'))}\left[\rho_{11}^+(m')\right]}, -0.5 \times \frac{\min_{\forall n'(n' \in (1,2,\ldots M'))}\left[C_{11}^+(n')\right]}{\max_{\forall n'(n' \in (1,2,\ldots M'))}\left[C_{11}^+(n')\right]}\right\}, \quad (23)$$

$$Subject\ to: \min_{\forall i(i=1,2,\ldots M)}\{\rho_{22}(i)\} > 0, \quad (24)$$

$$\min_{\forall i(i=1,2,\ldots M)}\{C_{22}(i)\} > 0, \quad (25)$$

**and Eqs. (15)-(17),**

where $M'$ ($M'< M$) is the first $M'$ sampling frequencies in the target operating frequency range ($f_{min}$, $f_{max}$) with $M$ sampling frequencies; $D'$ is the objective value; $ND'$ is the number of the sampling frequencies with the simultaneous negative $\rho_{11}$ and $C_{11}$ within the first $M'$ sampling frequencies; $m'\le M'$ and $n'\le M'$ are the serial numbers of the sampling frequencies with the positive $\rho_{11}$ and $C_{11}$, respectively; and $i\le M$ is the sampling frequency number. Briefly speaking, we will check the performance of the EMM at the first $M'$ sampling frequencies for the objective function in Eq. (23). However, the constraints in Eqs. (15)-(17), (24) and (25) are checked for all $M$ sampling frequencies. To avoid a too small search space, the values of $\delta_F$ and $\delta_C$ (see Eqs. (15) and (16)) are selected as 0.25 and 0.2, respectively. In order to guarantee that the longitudinal wave can transmit along the ΓY direction, we add two constraints, Eqs. (24) and (25), for all $M$ sampling frequencies. In this sense, if we produce a double-negative region within a frequency range, it is likely to come with a certain range of single-negativity where $\rho_{11}$ is positive and $C_{11}$ is negative. Then, the desired hyperbolic dispersion is obtained.

In fact, the above optimization setting is aiming at designing the double-negative dispersion in the first $M'$ sampling frequencies and the hyperbolic dispersion at the other $M-M'$ sampling frequencies. In calculation, we take $M'=9$, $M=11$, and ($f_{min}$, $f_{max}$)=(0.5 Hz, 23 kHz). Figure 16 presents the optimized lattice EMM. Unlike the EMM in Fig. 4(a), the structure in Fig. 16 shows stronger anisotropy with a lower symmetry. It has a horizontal big solid lump in the middle of the unitcell, which can seriously impact the vibration along *y*-direction. On one hand, the similar topological feature as shown in Fig. 4(a) can make sure the occurrence of the simultaneous negative $\rho_{11}$ and $C_{11}$. On the other hand, the horizontal solid lump prevents the EMM from producing the negative $\rho_{22}$ and $C_{22}$. Certainly, the double-negative range is decreased at the same time.

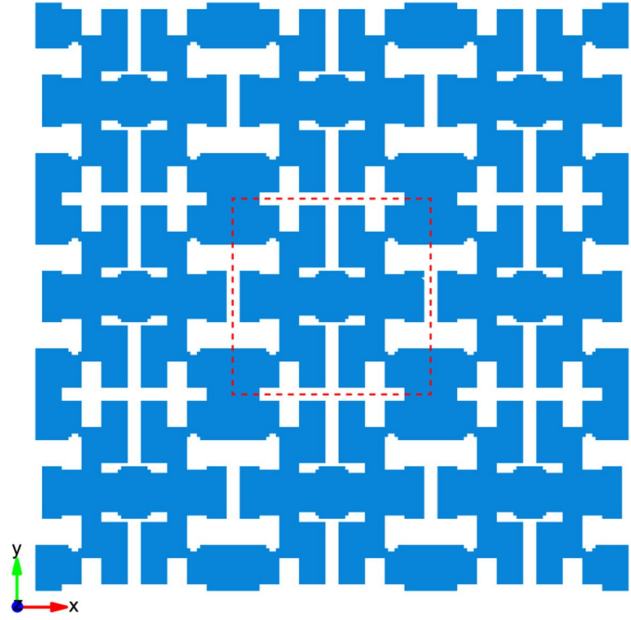

**Fig. 16.** Optimized super-anisotropic EMM ($f_{min}$=0.5 Hz, $f_{max}$=23 kHz, $M'$=9, $M$=11 and $\delta_C$=0.2).

We display the band structure, effective parameters, eigenstates and group velocities along two principle directions in Fig. 17. We can find obvious anisotropy from the band structure shown in Fig. 17(a), i.e., the distinguishing difference of the dispersion relations along ΓX and ΓY directions. The normalized quantity $q_x$ clearly shows the longitudinal and transverse wave motions along two directions. The forth band along ΓX direction is marked by the region L-x (16.28-18.68 kHz). The frequency range above the frequency of Γ point along ΓY direction is named L-y (18.68-22.81 kHz). According to the values of $q_x$, it is easy to find that these two regions only support the longitudinal waves along ΓX and ΓY directions, respectively. It is seen from Figs. 17(b) and 17(c) that both the mass densities and longitudinal moduli show strong anisotropy. The negative $\rho_{11}$ occurs in the range of 16.31-19.04 kHz. However, $\rho_{22}$ are keeping positive values in the range of 0.5 Hz-23.06 kHz. Besides, $C_{11}$ becomes negative in the range of 17.31-39 kHz, while $C_{22}$ is positive in the range of 0.5-38.28 kHz. Therefore, the double-negative index is generated in the range of 17.31-19.04 kHz; and the single-negative dispersion is produced (i.e., negative $C_{11}$, positive $\rho_{11}$, $\rho_{22}$ and $C_{22}$) in the range of 19.04-23.06 kHz. Moreover, $C_{12}$ is always positive in the range of 0.5 Hz-23 kHz. This means that the optimized EMM excellently avoids the transverse motions in the optimization frequency range. These features are conducive to both the purely longitudinal wave propagation along $x$- and $y$-directions.

To understand physics of the super-anisotropic dispersion, we present the four representative eigenstates in Fig. 17(a), see Fig. 17(d). Like mode $M_1$ in Fig. 6(d), eigenstate $E_1$ shows the emergence of negative $\rho_{11}$ and the similar quadrupolar-based resonance. The main difference is that the horizontal lump weakens the resonance to some extent, reducing the range of the negative $\rho_{11}$. Eigenstate $E_2$, which is obtained from the degenerate point of the forth band, has the similar resonance with eigenstate $E_1$. The rotation of the four corner solid blocks drives the horizontal lump to vibrate. Interestingly, the boundary vibration of eigenstate $E_2$ can be regarded as the shear deformation along ΓY direction. This property explains the corresponding transverse wave motion suggested by $q_x$ in Fig. 17(a). It is noted that eigenstate $E_3$ also shows the multipolar resonance. The bending vibration of the horizontal lump dominates the motion of the unitcell. On account of the insufficient bending ability, the optimized EMM can keep the positive $\rho_{22}$ in range L-y, Eigenstate $E_4$ clearly demonstrates the physical mechanism for the negative $C_{11}$, which is similar to mode $M_4$ in Fig. 9. Obviously, the existence of the horizontal lump also prevents the formation of the negative $C_{11}$. As a consequence, the EMM in Fig. 16 gets a narrower double-negative range compared with the EMM with a higher symmetry.

Furthermore, we present the group velocities of the forth band along ΓX and ΓY directions in Fig. 17(e). We can observe obvious negative $V_g$ along ΓX direction (16.28-18.68 kHz). However, $V_g$ along ΓY direction shows more complex property: it is negative in the range of 17.47-18.68 kHz and positive in 17.45-22.81 kHz. This is due to the fact that two eigenstates exist at the same frequency, which are caused by the longitudinal wave motion and quasi-transverse wave motion along ΓY direction, respectively. On the other hand, above the frequency of 18.68 kHz, only the positive $V_g$ along ΓY direction appears in the fourth band. Thus, in the corresponding frequency range, the longitudinal wave can only propagate along ΓY direction.

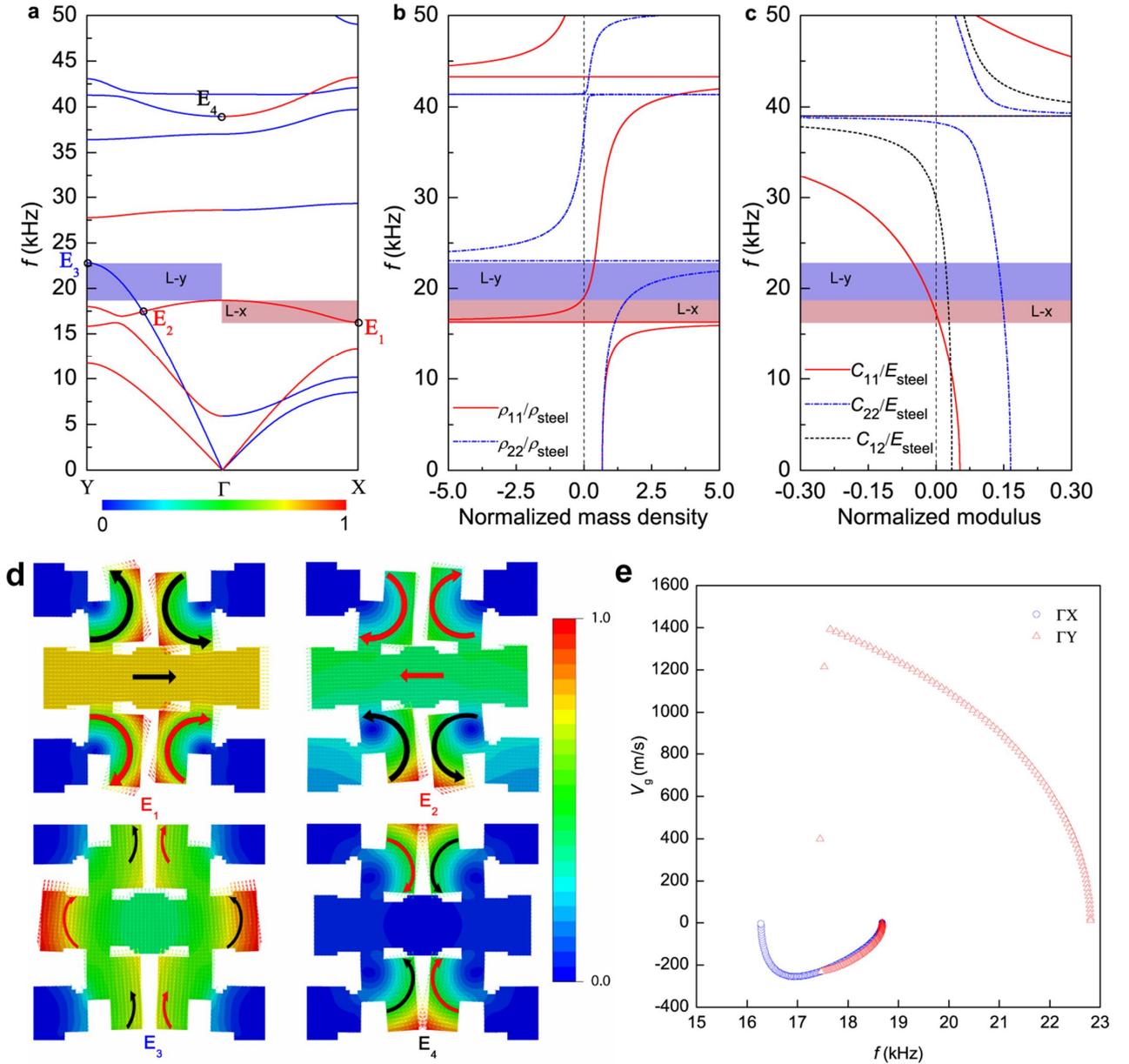

**Fig. 17.** Dispersion curves, effective material parameters and special eigenstates for the EMM in Fig. 16. **(a)** The band structure along ΓX and ΓY directions for the in-plane waves (the normalized quantity $q_x$ associated with the displacement

$u_x$ component is displayed by the corresponding color); **(b)** effective mass density along *x* (red solid line) and *y* (blue dash line) directions; **(c)** effective elastic moduli ($C_{11}$: red solid line; $C_{22}$: short dash dot line and $C_{12}$: short dash line); **(d)** field distributions of the eigenstates shown in (a); and **(e)** the group velocities of the forth band in (a) along ΓX (hollow circle) and ΓY (hollow triangle) directions. The frequency ranges for the longitudinal wave of the forth band in (a) along ΓX (L-x) and ΓY (L-y) directions are marked by the pink and blue bars, respectively.

To further confirm the super-anisotropic dispersion in Fig. 17, we present the corresponding EFCs, the imaging along *x*- and *y*-directions and their intensity profiles in Fig. 18. Figure 18(a) shows that the optimized EMM in Fig. 16 has double-negative dispersion long ΓX direction and hyperbolic dispersion along ΓY direction in the different frequency ranges. This is consistent with the results in Figs. 17(a)-17(c). Figures 18(b) and 18(c) show the imaging field patterns of the longitudinal waves at 17.7 and 22 kHz, respectively. Their corresponding wavelengths are defined as $\lambda_1$ and $\lambda_2$, respectively. Surprisingly, we observe that clear subwavelength imaging phenomena are produced for both *x*- and *y*-directions. Their intensity profiles yield the subwavelength resolution, FWHM=0.4$\lambda_1$ and FWHM=0.33$\lambda_2$. The gap between the group velocities (-207 and 615 m/s) at two frequencies also coincides with the resolution difference. Because a larger magnitude of the group velocity trends to qualitatively yield a higher imaging resolution, we argue that the hyperbolic mechanism in Figs. 17 and 18 can achieve higher resolution imaging than the double-negative one. Intuitively we believe this is due to the strong anisotropy which is very effective to reduce more energy loss. And yet the image generated by the hyperbolic EMM is very close to the right side of the EMM slab. This will challenge the image capture technology. But the double-negative EMM has more possibilities to realize the imaging in both near and far fields, see Figs. 13(b), 18(b) and 14(b). Anyhow, compared with the double-negative EMM, the optimized super-anisotropic EMM with two mechanisms of the double-negative and hyperbolic dispersions provides us more space for the subwavelength imaging.

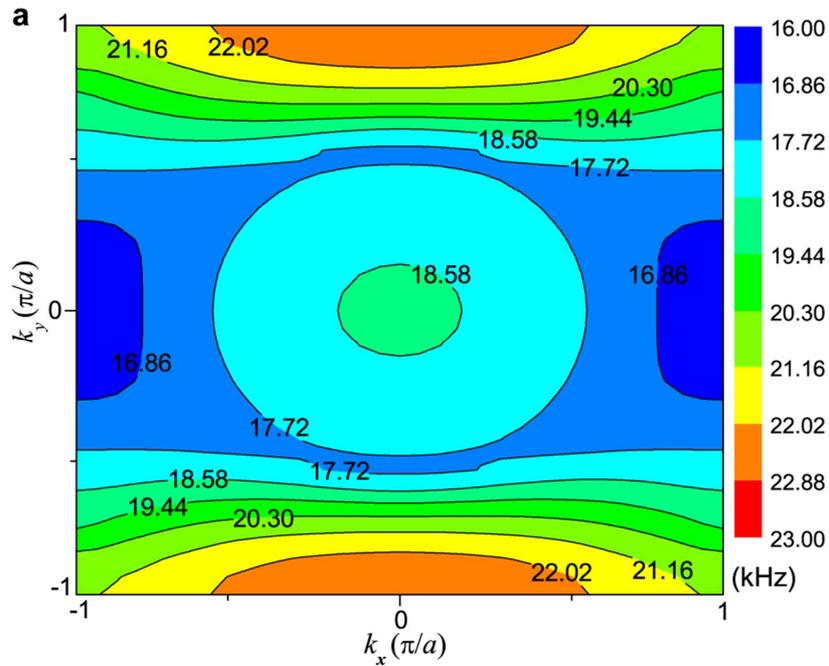

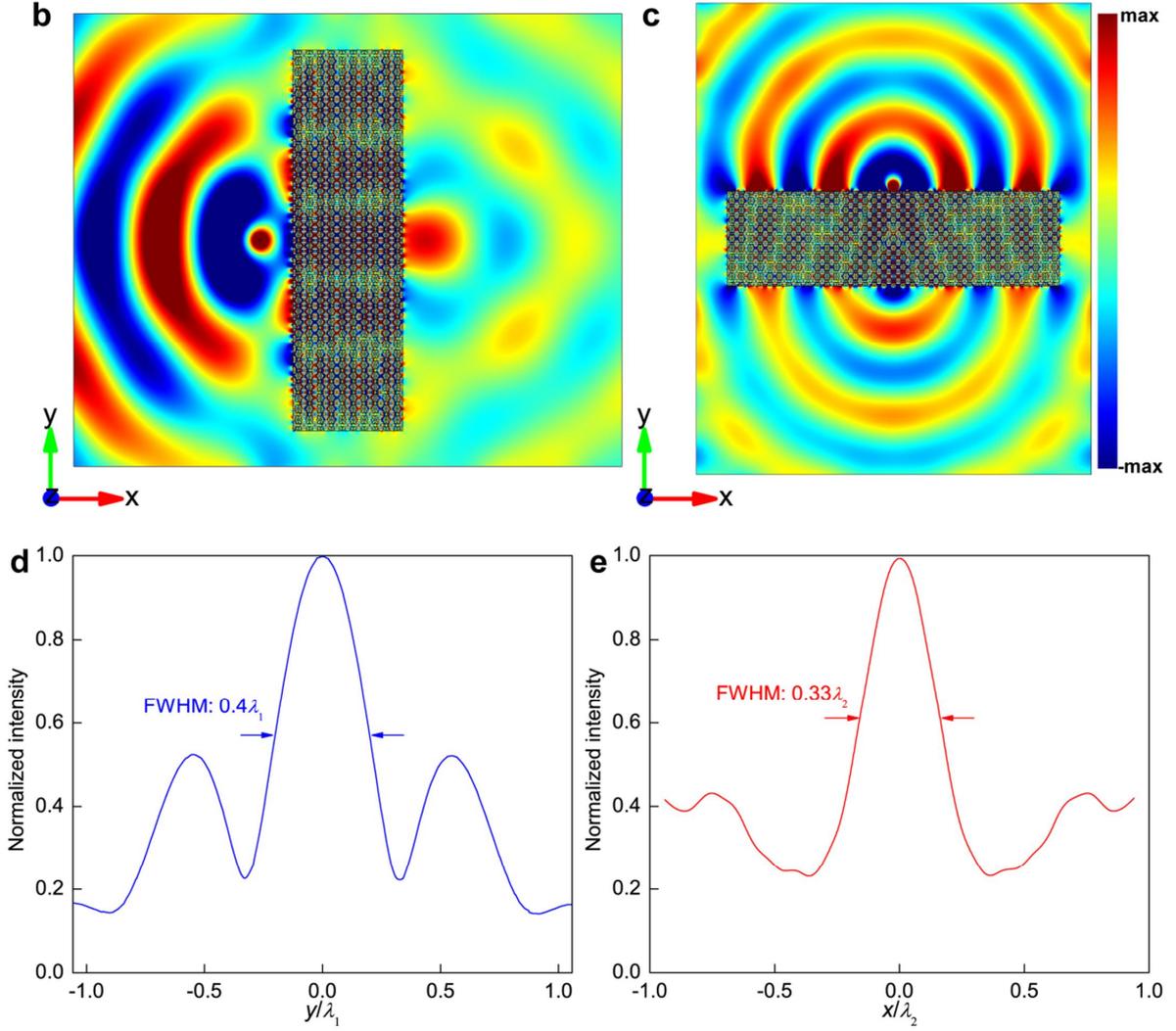

**Fig. 18.** EFCs and imaging of the EMM in Fig. 16. **(a)** The EFCs of the forth band in Fig. 17(a); **(b)** the imaging filed patterns of longitudinal wave at 17.7 kHz ($\lambda_1$) along *x*-direction; **(c)** the imaging filed patterns of longitudinal wave at 22 kHz ($\lambda_2$) along *y*-direction; **(d)** the intensity profile of image (FWHM=0.4$\lambda_1$) in (b); and **(e)** the intensity profile of image (FWHM=0.33$\lambda_2$) in (c).

## 6. Conclusions

In this paper, we have implemented evolutionary topology optimization of 2D anisotropic EMMs with the simultaneous negative mass density and negative longitudinal modulus within the desired operating frequency range along the prescribed direction. The metamaterial can be characterized by the effective medium theory which is agreed well with the dispersion relations. The contributions of the paper may be summarized as follows.

First, to obtain the broadband double-negative parameters, we propose an optimization function for maximizing difference between the minimal and maximal positive effective parameters at the sampling frequencies to design the orthotropic microstructure. Combined with the appropriate constraints, the proposed objective function is able to effectively find out the designs with double negativity in fairly broad subwavelength frequency regimes. The proposed optimization method shows robustness for the double-negative EMM design and is expected to be extended to other negative-index metamaterials.

Secondly, we present the optimized double-negative EMMs within different target frequency ranges. Especially, the optimized structure under ultra-low frequencies is promising in the fields of low-frequency insulation and vibration. All optimized anisotropic EMMs are composed of solid blocks and narrow connections, showing the typical local resonance features (Dong et al., 2014c). Through the mode analysis, we demonstrate that the double negativity is induced by the new quadrupolar or multipolar resonances which have not yet been discovered in EMMs due to limitation of artificial designs. This new mechanism will broaden the design space for negative-index EMMs.

Thirdly, the negative refraction and imaging of a longitudinal wave are realized by the optimized EMMs in the subwavelength scale. In addition to the negative longitudinal bands, most optimized orthogonal symmetrical EMMs exhibit the negative bands supporting only transverse waves as well. The negative refraction of a transverse wave in the single-phase EMM is observed for the first time. Note that if we employ a prism in our optimized EMMs, the similar wave mode conversion (Wu et al., 2011; Zhu et al., 2014) can emerge in the oblique interface as well. Moreover, for both longitudinal and transverse waves, the optimized double-negative EMM can serve as an anisotropic zero-index metamaterial along *x*-direction. The corresponding cloaking effects are demonstrated as well.

Finally, we perform the optimization with the more severe constraints and design a super-anisotropic EMM having double-negative and hyperbolic dispersions along two principle directions respectively. The proposed EMM produces simultaneous high-resolution images based on both double-negative and hyperbolic mechanisms, showing more possibilities for the elastic subwavelength imaging.

It is expected that our optimized EMMs with the unusual and fantastic properties and mechanisms will motivate more new ideas and novel applications including the low-frequency vibration attenuation, flat lens and ultrasonography for elastic waves in the future.

## Acknowledgements

This work is supported by the National Natural Science Foundation of China (Grant No. 11532001) and the Chinese Scholarship Council (CSC) and the German Academic Exchange Service through the Sino-German Joint Research Program (PPP) 2014. H. W. Dong would like to thank Professor O. Sigmund (Technical University of Denmark, Denmark) for his far-sighted suggestion and meaningful discussions. H. W. Dong also thanks doctoral students R. E. Christiansen (Technical University of Denmark, Denmark) and B. Wu (Zhejiang University, P. R. China) for their helpful discussions.